\documentclass[conference]{IEEEtran}
\IEEEoverridecommandlockouts
\usepackage{makeidx}  
\usepackage{amssymb}
\usepackage{color}
\usepackage{graphicx}
\usepackage{subfigure}
\usepackage{url}
\usepackage{makecell}
\usepackage{cite}
\usepackage{listings}
\usepackage{paralist}
\let\subparagraph\paragraph
\usepackage{cite}
\usepackage{amsfonts, amsmath, amssymb}
\usepackage{epstopdf}
\usepackage{float}
\usepackage{enumerate}
\usepackage{color}
\usepackage{multirow}
\usepackage{verbatim}
\usepackage{stfloats}
\usepackage{xcolor}
\usepackage[utf8]{inputenc}
\usepackage{amsfonts}
\usepackage{eufrak}
\usepackage{eucal}
\usepackage{amsmath}
\usepackage{makeidx}
\usepackage{bm}
\usepackage{booktabs}
\usepackage{wrapfig}
\usepackage[font=small,skip=4pt]{caption}

\makeatletter
\newif\if@restonecol
\makeatother

\usepackage[ruled,vlined, linesnumbered]{algorithm2e}
\makeatletter
\def\footnoterule{\kern-3\p@
  \hrule \@width 2in \kern 2.0\p@} 
\makeatother

\newtheorem{mydef}{Definition}[section]

\makeatletter
\def\BState{\State\hskip-\ALG@thistlm}
\makeatother

\definecolor{codegreen}{rgb}{0,0.6,0}
\definecolor{codegray}{rgb}{0.5,0.5,0.5}
\definecolor{codepurple}{rgb}{0.58,0,0.82}
\definecolor{backcolour}{rgb}{0.95,0.95,0.92}

\definecolor{dkgreen}{rgb}{0,0.6,0}
\definecolor{gray}{rgb}{0.5,0.5,0.5}
\definecolor{mauve}{rgb}{0.58,0,0.82}
\lstdefinestyle{mystyle}{
  backgroundcolor=\color{backcolour},   commentstyle=\color{codegreen},
  keywordstyle=\color{magenta},
  numberstyle=\tiny\color{codegray},
  stringstyle=\color{codepurple},
  basicstyle=\footnotesize,
  breakatwhitespace=false,
  breaklines=true,
  keepspaces=true,
  numbers=left,
  numbersep=5pt,
  showspaces=false,
  showstringspaces=false,
  showtabs=false,
  tabsize=2
}

\makeatletter
\newcommand{\tabincell}[2]{\begin{tabular}{@{}#1@{}}#2\end{tabular}}
\newcommand{\xRightarrow}[2][]{\ext@arrow 0359\Rightarrowfill@{#1}{#2}}
\makeatother

\newcommand{\ccsl}{\textsc{Ccsl}}
\newcommand{\tdl}{\textsc{Tadl2}}

\newcommand{\gt}[1]{\texttt{#1}}

\newcommand{\uppaal}{\textsc{Uppaal}}

\newcommand{\ed}{\textsc{East-adl}}

\newcommand{\smc}{\textsc{Uppaal-SMC}}

\newcommand{\marte}{\textsc{MARTE}}

\newcommand{\protl}{{ProTL}}

\begin{document}
%
\title{Formal Verification of Dynamic and Stochastic Behaviors for Automotive Systems}
\author{
\IEEEauthorblockN{
Li Huang\IEEEauthorrefmark{1}, Tian Liang\IEEEauthorrefmark{1} and
Eun-Young Kang\IEEEauthorrefmark{2}
}
\IEEEauthorblockA{\IEEEauthorrefmark{1}School of Data \& Computer Science, Sun Yat-Sen University, Guangzhou, China \\
\{huangl223, liangt37\}@mail2.sysu.edu.cn}
\IEEEauthorblockA{\IEEEauthorrefmark{2}The Maersk Mc-Kinney Moller Institute, University of Southern Denmark, Denmark\\
eyk@mmmi.sdu.dk}
}
\maketitle

\begin{abstract}
Formal analysis of functional and non-functional requirements is crucial in automotive systems. The behaviors of those systems often rely on complex dynamics as well as on stochastic behaviors.
We have proposed a probabilistic extension of Clock Constraint Specification Language, called Pr\ccsl, for specification of (non)-functional requirements and proved the correctness of requirements by mapping the semantics of the specifications into \uppaal\ models. Previous work is extended in this paper by including an extension of Pr\ccsl, called Pr\ccsl$^*$, for specification of stochastic and dynamic system behaviors, as well as complex requirements related to multiple events.
To formally analyze the system behaviors/requirements specified in Pr\ccsl$^*$, the Pr\ccsl$^*$ specifications are translated into stochastic \uppaal\ models for formal verification. We implement an automatic translation tool, namely \protl, which can also perform formal analysis on Pr\ccsl$^*$ specifications using \smc\ as an analysis backend.
Our approach is demonstrated on two automotive systems case studies.
\end{abstract}



\section{Introduction}

\label{sec:introduction}
Model-based development (MBD) is rigorously applied in automotive systems in which the software controllers continuously interact with physical environments. The behaviors of automotive systems often involve complex hybrid dynamics as well as stochastic characteristics.
Formal verification and validation (V\&V) technologies are indispensable and highly recommended for development of safe and reliable automotive systems \cite{iso26262,iec61508}.
\emph{Statistical model checking (SMC)} techniques have been proposed \cite{smc-challenge,smc-david-12,david2015uppaal} to address the reliability of hybrid systems associated with the stochastic and non-linear dynamical features. These techniques for fully stochastic models validate probabilistic properties of controllers in given environments under uncertainty.

\ed\ \cite{EAST-ADL, maenad} is a concrete example of MBD approach for architectural modeling of automotive systems.
The latest release of \ed\ has adopted the time model proposed in the Timing Augmented Description Language (\tdl) \cite{TADL2}, which expresses and
composes the basic timing constraints, i.e., repetition rates, end-to-end delays, and synchronization constraints.
\tdl\ specializes the time model of \marte, the UML profile for Modeling and Analysis of Real-Time and Embedded systems \cite{MARTE}.
\marte\ provides \ccsl\ \cite{andre2009syntax, ccslcorrectness}, which is the clock constraint specification language for specification of temporal constraints and functional causality properties \cite{khan2016combining}.
In \ccsl, time can be either chronometric (i.e., associated with physical time) or logical (i.e., related to events occurrences), which are represented by dense clocks and logical clocks, respectively.
Dense clocks can be multiform and attached with various rates, allowing to express the evolution of time-related quantities with different units, e.g., temperature degree and crankshaft angle \cite{peraldi2012timing}.
The discrete and dense  clocks in  \ccsl\ enable the specifications of hybrid system behaviors that incorporate both discrete phenomena and continuous dynamics \cite{liu2013hybrid}.

We have previously proposed a probabilistic extension of \ccsl, i.e., Pr\ccsl\ \cite{ifm18, fase2019}, to formally specify (non)-functional properties in weakly-hard (WH) context \cite{Bernat2001Weakly}, i.e., a bounded number of constraints violations would not lead to system failures when the results of the violations are negligible.
However, Pr\ccsl\ still lacks expressivity for describing critical system behaviors regarding to: \begin{inparaenum}
\item Continuous dynamic behaviors of physical plant, which are typically modeled by ordinary differential equations (ODE) containing functions and derivatives;
\item Discontinuous activities triggered by discrete events.  For example, the velocity of an automotive system undertakes instantaneous changes when collisions occur;
\item Stochastic time spans of continuous activities (e.g., execution time of a component, response time for a spontaneous failure);
\item Nondeterministic behaviors regarding to exclusive activities. For instance, during the execution of an automotive system, multiple sensors can forward messages to the controller simultaneously and the order for the controller to receive the messages is nondeterministic.
\end{inparaenum}

Supporting formal specifications of dynamic and stochastic behaviors is crucial for effective analysis of automotive systems.
In this paper, we propose an extension of Pr\ccsl, called Pr\ccsl$^*$, to enable the specification of the aforementioned system behaviors.
To allow the analysis of system behaviors specified in Pr\ccsl$^*$, the Pr\ccsl$^*$ specifications are translated into \smc\ models for formal verification.
Furthermore, in Pr\ccsl, requirements are specified as binary clock constraints that describe temporal and causal relations between two events.
To support the analysis of complex requirements associated with multiple events, we extend the binary clock constraints into n-ary constraints and provide the corresponding translation patterns in \smc.
The automatic translation from Pr\ccsl$^*$ specifications (of system behaviors/requirements) to \smc\ models is implemented in a tool called \protl\ (Probabilistic \ccsl\ Translator). Formal analysis of the Pr\ccsl$^*$ specifications can be performed by \protl\ using \smc\ as an analysis backend.
Our approach is demonstrated on two automotive systems case studies.

The paper is organized as follows: Sec. \ref{sec:preliminary}
presents an overview of Pr\ccsl\ and \smc.
The definition of Pr\ccsl$^*$ is presented in Sec. \ref{sec: extension}.
Sec. \ref{sec:translation} describes the mapping rules from Pr\ccsl$^*$ specifications into stochastic \uppaal\ models.
 The applicability of our approach is demonstrated by performing verification on two automotive systems in Sec. \ref{sec:case-study}. Sec. \ref{sec:r-work} and  Sec. \ref{sec:conclusion} present related works and conclusion.

\section{Preliminary}
\label{sec:preliminary}
In this section, we first introduce the notations and formalism in Pr\ccsl\ that are employed in the rest of the paper.
Then, we give an overview of \smc, which is utilized as the analysis backend of \protl.

\noindent\textbf{PrCCSL} \cite{ifm18} is a probabilistic extension of \ccsl\ (Clock Constraint Specification Language)  \cite{andre2009syntax}, which specifies temporal and causal constraints with stochastic characteristics in weakly-hard context \cite{Bernat2001Weakly}.
In Pr\ccsl, a clock is a basic element that represents a sequence of (possibly infinite) instants.
A clock in Pr\ccsl\ can be either dense or discrete/logical. A dense clock represents physical time, which is considered as a continuous and unbounded progression of time instants.
The physical time is represented by a dense clock with a base unit.
\emph{idealClk} is a predefined dense clock in \ccsl, which represents the ideal physical time whose unit is second.
A dense clock can be discretized into a discrete/logical clock.
For example, a clock $ms$ can be defined based on $idealClock$: $ms$ = $idealClock$ {\gt{discretizedBy}} 0.001, i.e., $ms$ ticks every 1 millisecond.
A logical clock represents an \emph{event} and the clock instants correspond to the event occurrences.
A clock represents an instance of {\small \gt{ClockType}} characterized by a set of attributes. The keyword {\small \gt{DenseClockType}} ({\small \gt{DiscreteClockType}}) defines new dense (discrete) clock types.

\begin{table*}[b]
\centering
\caption{Clock Expressions}
\begin{tabular}{llp{350pt}}
\toprule  %
Expression& Notation &  Remarks \\
\midrule  %
\tabincell{l}{ITE\\
(if-then-else)}
& \tabincell{l}{$res \triangleq\ b$ \textcolor{red}{?} $c1$ \textcolor{red}{:} $c2$}
& \tabincell{l}{$res$ behaves either as $c1$ or as $c2$  based on the value of  the boolean variable $b$.}
\\
\hline

\tabincell{l}{DelayFor}
& \tabincell{l}{$res \triangleq\ ref$ ($d$) \textcolor{red}{$\rightsquigarrow$} $base$}
&\tabincell{l}{ $\forall i \in$ $\mathbb{N^{+}}$, the $i^{th}$ tick of $res$ corresponds to the $i^{th}$  tick of $ref$ delayed for $d$ ticks (or units) of $base$.}
\\
\hline
\tabincell{l}{FilterBy}
& \tabincell{l}{$res \triangleq\ {base}\ \textcolor{red}{\blacktriangledown}\ u(v)$}
& \tabincell{l}{$res$ is generated by filtering the instants of \emph{base} based on a binary word \emph{w=u(v)}, where $u$ is  \emph{prefix} and $v$ is\\ \emph{period}. ``$(v)$''  denotes the infinite repetition of $v$. $\forall i \in$ $\mathbb{N^{+}}$, if the $i^{th}$ bit of $w$  is 1, $res$ ticks at the $i^{th}$ tick\\ of $base$.}
\\
\hline
\tabincell{l}{PeriodicOn}
& \tabincell{l}{$res \triangleq\ base\ \textcolor{red}{\propto}\ n$}
& \tabincell{l}{Any two consecutive instants of $res$ are separated  by $n$ instants of  \emph{base}.}
\\
\hline
\tabincell{l}{Infimum}
& \tabincell{l}{$res\ \triangleq\ c1\ \textcolor{red}{\wedge}\ c2$}
& \tabincell{l}{$res$ is the slowest clock faster than $c1$ and $c2$}.
\\
\hline
\tabincell{l}{Supremum}
& \tabincell{l}{$res\ \triangleq\ c1\ \textcolor{red}{\vee}\ c2$}
& \tabincell{l}{$res$ is the fastest clock slower than $c1$ and $c2$}.
\\
\hline
\end{tabular}
  \label{expression}%
\end{table*}

Pr\ccsl\ provides two types of clock constraints to describe the occurrences of different events (logical clocks), i.e., clock \emph{expressions} and clock \emph{relations}. \emph{Expressions} derive new clocks from the already defined clocks. Table \ref{expression} presents a set of clock \emph{expressions}, where $c1, c2, ref, res, base$ are clocks and ``$\triangleq$'' means ``is defined as''.
We only list a subset of clock \emph{expressions} here due to page limit and the full set of \emph{expressions} can be found in \cite{andre2009syntax}.

A clock \emph{relation} limits the occurrences among different clocks/events,
which are defined based on {\small \gt{run}} and {\small \gt{history}}.
A {\small \gt{run}} corresponds to an execution of the system model
where the clocks tick/progress. The {\small \gt{history}} of a clock $c$ represents the number of times $c$ has ticked currently.
A probabilistic \emph{relation} in Pr\ccsl\ is satisfied if and only if the probability of
the \emph{relation} constraint being satisfied is greater than or equal
 to the probability threshold \emph{p} $\in [0, 1]$.
 Given $k$ {\gt{runs}} $=$ $\{R_1, \ldots, R_k \}$, the probabilistic \emph{relations} in Pr\ccsl, including {\small \gt{subclock}},
  {\small \gt{coincidence}}, {\small \gt{exclusion}},  {\small \gt{precedence}} and {\small \gt{causality}} are defined in Table \ref{relation}.

\begin{table*}[t]
\centering
\caption{Relations in Pr\ccsl}
\begin{tabular}{lll}
\toprule  %
Relation& Notation &  Remarks \\
\midrule  %
\tabincell{l}{Probabilistic \\Subclock}
& \tabincell{l}{$c1 \textcolor{red}{\subseteq_p} c2$}
& \tabincell{l}{$c1 \textcolor{red}{\subseteq_p} c2 \Longleftrightarrow
Pr[c1 \textcolor{red}{\subseteq} c2] \geqslant p$, where
$Pr[c1\textcolor{red}{\subseteq} c2]  = \frac{1}{k} \sum\limits_{j=1}^{k} \{R_j \models c1 \textcolor{red}{\subseteq}
 c2\}$ is the ratio of {\gt{runs}} that satisfies the relation \\out of k {\gt{runs}}.
A {\gt{run}} $R_j$ satisfies the {\gt{subclock}} relation $c1 \textcolor{red}{\subseteq} c2$
if ``when $c1$ ticks, $c2$ must tick'' always holds in $R_j$.
}
\\
\hline

\tabincell{l}{Probabilistic \\ Coincidence}
& \tabincell{l}{$c1 \textcolor{red}{\equiv_p} c2$}
&\tabincell{l}{ $c1 \textcolor{red}{\equiv_p} c2 \Longleftrightarrow
 Pr[c1 \textcolor{red}{\equiv} c2] \geqslant p$, where $Pr[c1\textcolor{red}{\equiv} c2]  = \frac{1}{k}
 \sum\limits_{j=1}^{k} \{R_j \models c1 \textcolor{red}{\equiv} c2\}$ is the ratio of {\gt{runs}}
 that satisfies the coinci-\\dence relation out of k {\gt{runs}}. The satisfaction of {\gt{coincidence}} relation $c1 \textcolor{red}{\equiv} c2$ is established in $R_j$ when the two \\conditions ``if $c1$ ticks, $c2$ must tick'' and ``if $c2$ ticks, $c1$ must tick'' always hold.}
\\
\hline
\tabincell{l}{Probabilistic \\ Exclusion}
& \tabincell{l}{$c1 \textcolor{red}{\#_p} c2$}
& \tabincell{l}{$c1 \textcolor{red}{\#_p} c2 \Longleftrightarrow Pr[c1 \textcolor{red}{\#} c2] \geqslant p$, where $Pr[c1\textcolor{red}{\#} c2]  = \frac{1}{k} \sum\limits_{j=1}^{k} \{R_j \models c1 \textcolor{red}{\#} c2\}$, indicating the ratio of {\gt{runs}} that satisfies the \\exclusion relation out of k {\gt{runs}}. A {\gt{run}} $R_j$ satisfies the {\gt{exclusion}} relation $c1 \textcolor{red}{\#} c2$  if the condition ``$c1$ and $c2$ must \\not tick at the same time'' holds.}
\\
\hline
\tabincell{l}{Probabilistic \\Precedence}
& \tabincell{l}{$c1 \textcolor{red}{\prec_p} c2$}
& \tabincell{l}{$c1 \textcolor{red}{\prec_p} c2 \Longleftrightarrow Pr[c1 \textcolor{red}{\prec} c2] \geqslant p$,
where $Pr[c1\textcolor{red}{\prec} c2]  = \frac{1}{k} \sum\limits_{j=1}^{k} \{R_j \models c1 \textcolor{red}{\prec} c2\}$, which denotes the ratio of {\gt{runs}} that satisfies\\ the precedence relation out of k {\gt{runs}}.
A {\gt{run}} satisfies the {\gt{precedence}} relation $c1 \textcolor{red}{\prec} c2$ if the history of $c1$ is not less \\than the history of $c2$,
and $c2$ must not tick when their histories are equal.}
\\
\hline
\tabincell{l}{Probabilistic \\Causality}
& \tabincell{l}{$c1 \textcolor{red}{\preceq_p} c2 $}
& \tabincell{l}{$c1 \textcolor{red}{\preceq_p} c2 \Longleftrightarrow Pr[c1 \textcolor{red}{\preceq} c2]
\geqslant p$, where $Pr[c1 \textcolor{red}{\preceq} c2]$  $=$ $\frac{1}{k}$
$\sum\limits_{j=1}^{k} \left\{R_j \models c1 \textcolor{red}{\preceq} c2 \right\}$,
 i.e., the ratio of {\gt{runs}} satisfying the {\gt{causa}}-\\{\gt{lity}} relation among
 the total number of $k$ {\gt{runs}}. A {\gt{run}} $R_j$ satisfies the {\gt{causality}}
 relation $c1 \textcolor{red}{\preceq} c2$ if the history of $c2$\\ is always less than or equal to the history of $c1$.
}
\\

\bottomrule %
\end{tabular}
  \label{relation}%
\end{table*}

\vspace{0.05in}
\noindent\textbf{UPPAAL-SMC} \cite{smc} performs the probabilistic analysis of properties by monitoring simulations of complex hybrid systems in a given stochastic environment and using results from the statistics to determine whether the system satisfies the property with some degree of confidence. Its clocks evolve with various rates, which are specified with \emph{ordinary differential equations} (ODE).
\smc\ provides a number of queries related to the stochastic interpretation
of Timed Automata (STA) \cite{david2015uppaal} and they are as follows, where $N$ and $bound$ indicate the number of simulations to be performed  and the time bound on the simulations respectively:
\begin{inparaenum}
\item \emph{Probability Estimation} estimates the probability of a requirement property $\phi$ being satisfied for a given STA model within the time bound: $Pr[bound]\ \phi$.
\item \emph{Hypothesis Testing} checks if the probability of $\phi$ being satisfied is larger than or equal to a certain probability $P_0$: $Pr[bound]\ \phi \ \geqslant\ P_0$.
\item \emph{Probability Comparison} compares the probabilities of two properties being satisfied in certain time bounds: $Pr[bound_1]$ $\phi_1$ $\geqslant$ $Pr[bound_2]$ $\phi_2$.
\item \emph{Expected Value} evaluates the minimal or maximal value of a clock or an integer value while \smc\ checks the STA model: $E[bound; N](min:\phi)$ or $E[bound; N](max:\phi)$.
\item \emph{Simulations}: \smc\ runs $N$ simulations on the STA model and monitors $k$ (state-based) properties/expressions $\phi_1,..., \phi_k$ along the simulations within simulation bound $bound$: $simulate$ $N$ $[\leqslant$ $bound]\{\phi_1,..., \phi_k\}$.
\end{inparaenum}

\section{Extension of PrCCSL}
\label{sec: extension}

To improve the expressiveness of Pr\ccsl\ for specifying dynamic and stochastic system behaviors (i.e., the four types of behaviors mentioned in Sec. \ref{sec:introduction}), we present an extension of  Pr\ccsl, called Pr\ccsl$^*$, which is augmented with notations of interval-based {\small \gt{DelayFor}},  {\small \gt{Clock Action}},  {\small \gt{Probabilistic Clock Action}} and {\small \gt{DenseClockType}}.
Furthermore, to allow specification of complex requirements involving multiple events, the binary clock \emph{relations} in Pr\ccsl\ are extended into n-ary clock \emph{relations}.

In automotive systems, time delays between events/activities (e.g., message transmissions, execution of computational components) can be stochastic and fluctuate randomly in the environments under uncertainty. Those stochastic time delays can be described as random variables that uniformly distributed between the values of the \emph{best-case time} and the \emph{worst-case time}.
To express the stochastic time delay between events, we define an interval-based {\small \gt{DelayFor}} \emph{expression}.

\begin{mydef}[{{Interval-based DelayFor}}]
Let $C$ be a set of clocks. $res$, $ref$, $base$ $\in$ $C$, $lower$, $upper$ $\in$ $N^{+}$ and $upper$ $\geq$ $lower$, the interval-based {\small\gt{DelayFor}} generates a new clock $res$ by delaying $ref$ for random number of units (instants) of $base$ clock, which is expressed as:
\[res\ \triangleq\ ref\ ([lower, upper])\ \textcolor{red}{\rightsquigarrow}\ base\]
\noindent where $lower$, $upper$ are two positive integers that represent the minimum and maximum time delay. The {expression} is interpreted as: $res \triangleq\ ref (x)\ \textcolor{red}{\rightsquigarrow}\ base \ \wedge$  $x \in$ $[lower$, $upper]$, i.e., $res$ is a clock generated by delaying $ref$ for $x$ units of $base$ and $x$ is given by (continuous/discrete) uniform probability distribution over $[lower$, $upper]$.
\end{mydef}

The standard {\small\gt{DelayFor}} operator (see Table \ref{expression}) can be seen as a special interval-based {\small\gt{DelayFor}} when $lower=upper$.
Since the $base$ clock of the {\small\gt{DelayFor}} operator can be either dense or discrete, the delay should conform to continuous or discrete probability distribution according to the clock type of $base$.

Automotive systems are event-driven systems that react to external/internal events, e.g., triggering of sensors/actuators or arrivals of input signals. Event occurrences can lead to execution of related actions, i.e., functions/operations on variables.
The ``event-action'' relation can be specified by {\small \gt{Clock Action}}.

\begin{mydef}[{{Clock Action}}]
Let $V$ be a set of variables and $C$ be a set of logical clocks. $Assign$ denotes a sequence of assignments to $V$ and $Func$ represents a set of mathematical functions on $V$.
{\small \gt{Clock action}} is a function $A$: $C$ $\mapsto$ $Assign \cup Func$. Let $c$ $\in$ $C$, $A(c)$ represents the set of assignments and functions that are invoked to execute when $c$ ticks. The {\small \gt{clock action}} of $c$ is defined as:
\[c\ {\textcolor{red}{\rightarrow}}\ \{\lambda_1,\ \lambda_2,\ ...,\ \lambda_n\}\]
\noindent where $n \in \mathbb{N}^{+}$ is the number of assignments/functions related to $c$, $\lambda_i$ $\in$ $Assign \cup Func$ ($i \in \{1, 2, ..., n\}$) denotes an assignment or a function that is executed when an occurrence of $c$ is detected.
\end{mydef}

In other words, the {\small \gt{clock action}} relates a logical clock $c$ with a set of operations that are performed to change the system behaviors. Note that the assignment of each variable in $V$ must appear at most once in the {\small \gt{clock action}}, i.e.,  two or more assignments to the same variable are not allowed.

An event can be associated with multiple actions under uncertainty, i.e., it is uncertain which action out of a set of actions the system will take when the event occurs.
In this case, the actions related to the same event can be assigned with probabilities and interpreted as probabilistic alternatives. We enrich {\small\gt{clock actions}}  with probabilities and define {\small\gt{probabilistic clock action}}.

\begin{mydef}[{{Probabilistic Clock Action}}]
Let $C$ be a set of logical clocks and $P$ be a set of real numbers in [0, 1]. The {\small \gt{probabilistic clock action}} is a function $A_p$: $C$ $\times$ $Assign \cup Func$ $\mapsto$ $P$. Let $c$ $\in$ $C$ and $\Lambda_i$ $\subseteq$ $Assign \cup Func$ ($i \in \{1, 2, ..., n\}$),  $A_p(c, \Lambda_i)$ represents the probability of $\Lambda_i$ being executed when $c$ ticks. The probabilistic clock action is represented as:
\[c\ {\textcolor{red}{\rightarrow_p}}\ \{(p_1, \Lambda_1),\ (p_2, \Lambda_2),\ ...,\ (p_n, \Lambda_n)\}\]
\noindent where $n \in \mathbb{N}^{+}$ is the number of sets of actions related to $c$, $\Lambda_i$ is a set of assignments/functions, and $p_i \in P$ represents the probability of actions in $\Lambda_i$ being executed when $c$ occurs, i.e., $p_i=Pr(A(c)=\Lambda_i)$ and $\sum\limits_{i=1}^{n} p_i=1$.
\end{mydef}

We express the probability distribution of the set of actions related to event $c$ by a list of tuples in the form of ``($p_i$, $\Lambda_i$)''. For instance, $c$ \textcolor{red}{$\rightarrow_p$} \{(0.2, \{$v=0$\}), (0.3, \{$v=1$\}), (0.5, \{$v=2$\})\} means that when $c$ ticks, $v$ is assigned the value 0, 1, 2 with probability 0.2, 0.3 and 0.5, respectively.

In automotive systems, variations of physical quantities (e.g., energy consumption, temperature) usually involve continuous
dynamics described by ODE, as well as discrete changes activated by physical phenomena. For example, battery consumption of an automotive system increases continuously with a certain rate when the vehicle runs under certain modes (e.g., braking or turning) while undergoes discrete increments when the physical phenomena (e.g., turning on/off switches in circuits) take place. To allow the specification of the continuous/discrete variations of those physical quantities, we utilize dense clocks to represent those quantities and extend the attributes of {\small \gt{DenseClockType}}, from which the dense clocks can be instantiated.

\begin{mydef}[{{DenseClockType}}]
Let $n$ and $m$ be two positive integers. A dense clock type $DT$ can be defined based on the following four attributes:

\noindent {\small\gt{DenseClockType}} $DT$
\{{\small\gt{reference}} $ref$, {\small\gt{factor}} $r$,
{\small\gt{offset}} \{($c_1$, $v_1$), …, ($c_n, v_n$)\},
{\small\gt{reset}} \{$e_1$,\ …,\ $e_m$\}\};
\vspace{0.05in}

\noindent where
\begin{itemize}
\item[--] {\small\gt{reference}} specifies a referential dense clock, i.e., $ref$.

\item[--] {\small\gt{factor}} indicates the increase rate of instances of $DT$ compared to $ref$.

\item[--] {\small\gt{offset}} represents a set of tuples \{($c_1$,\ $v_1$),\ ($c_2$,\ $v_2$) …,\ ($c_n$,\ $v_n$)\}, where $\forall i \in \{1,\ 2,\ ...,\ n\}$,  $c_i \in C$ is a logical clock and $v_i\in \mathbb{R}$ represents a real number.
    The ticks of $c_i$ result in the instaneous changes of instances of $DT$ by $v_i$ time units.

\item[--] {\small\gt{reset}} represents a set of logical clocks \{$e_1$,\ $e_2$\ …,\ $e_m$\} whose ticks reset the instances of $DT$.
\end{itemize}
\end{mydef}

Let $c$ be an instance of $DT$ and $v_c$ be a real number that represents the time value of $c$.
The {\small\gt{factor}} of $c$ (denoted by $r$) corresponds to the increase rate of $v_c$ compared to the {\small\gt{reference}} clock.
For example, if the {\small\gt{reference}} clock of $c$ is $idealClk$,
then $v_c=\int r \ dt$, where $t\in[0,\ \infty]$ represents the physical time.
The {\small\gt{offset}} and  {\small\gt{reset}} of $c$ describe discrete changes of $v_c$ triggered by
events, i.e., $\forall i \in \{1,\ 2,\ ...,\ n\},\ c_i$ \textcolor{red}{$\rightarrow$} \{$v_c=v_c+v_i$\} and $\forall\ j \in \{1,\ 2,\ ...,\ m\}$, $e_j$ \textcolor{red}{$\rightarrow$} \{$v_c=0$\}.

By utilizing the operators and notations in Pr\ccsl$^*$, a system can be specified as a {\small \gt{Probabilistic Clock Based System (PCBS)}}, in which the continuous and discrete system behaviors can be described as the evolutions of a set of dense and logical clocks.

\begin{mydef}[{{Probabilistic Clock Based System (PCBS)}}]
{A {\small\gt{probabilistic clock based system}} is a tuple}:
\[S\ =\ \langle\ T,\ C_t,\ C_n,\ Exp,\ A,\ A_p\ \rangle\]
\noindent {where}

\begin{itemize}

\item[--] {$T$ is a set of clock types};

\item[--] {$C_t$ is a set of dense clocks};

\item[--] {$C_n$ is a set of logical clocks};

\item[--] {$Exp$ is a set of clock expressions};

\item[--] {$A$ is a set of {\small\gt{clock actions}} of clocks in $C_n$};

\item[--] {$A_p$ is a set of {\small\gt{probabilistic clock actions}} of clocks in $C_n$}.
\end{itemize}
\end{mydef}

Clocks in $C_t$ and $C_n$ are instances derived from clock types in $T$.
$C_t$ and $C_n$ are two exclusive sets. {\small\gt{Clock actions}} and {\small\gt{probabilistic clock actions}} (see Definition 3.2 and 3.3) are employed to describe the actions activated by events (i.e., logical clocks).
Since clock \emph{relations} in Pr\ccsl$^*$ are applied in specifying requirements and not utilized in system behaviors specifications, a {\small\gt{PCBS}} does not contain any clock \emph{relations}.

In Pr\ccsl, requirements are specified by clock \emph{relations} (see Table \ref{relation}) between two events.
To describe the complex requirements associated with multiple events, we extend the binary \emph{relations} into n-ary \emph{relations}, which allow to describe the dependencies among a set of events.

\begin{mydef}[{{N-ary Relation}}]
Let $\mathcal{M}$ be a {\small\gt{PCBS}} and $c_1$, $c_2$, ..., $c_n$ are $n$ clocks in $\mathcal{M}$. An $n$-ary clock \emph{relation} among clocks $c_1$, $c_2$, ..., $c_n$, denoted as $\textcolor{red}{\sim}(c_1, c_2, ..., c_n)$, is satisfied over $\mathcal{M}$ if the following condition holds:

\[\mathcal{M} \vDash \textcolor{red}{\sim}(c_1, c_2, ..., c_n) \Longleftrightarrow\ \forall i, j: 1 \leq i < j \leq n \Rightarrow \mathcal{M} \vDash c_i \textcolor{red}{\sim} c_j\]
\noindent where \textcolor{red}{$\sim$} $\in \{\textcolor{red}{\subseteq, \equiv, \prec, \preceq, \#} \}$, $n \geq 2$ is the number of clocks in the \emph{relation}.
\end{mydef}

Note that the $n$ clocks in the n-ary \emph{relations} are partially ordered,
e.g., $\textcolor{red}{\subseteq} (c_1,c_2,c_3)$ and $\textcolor{red}{\subseteq} (c_2, c_1, c_3)$ are different {\small\gt{subclock}} \emph{relations}.
Informally, an n-ary \emph{relation}, including {\small\gt{coincidence}}, {\small\gt{subclock}}, {\small\gt{exclusion}}, {\small\gt{precedence}} and {\small\gt{causality}} \emph{relations}, is satisfied if any order-preserving pair (i.e., the order relation $i<j$ is maintained) of two clocks in the set of $n$ clocks satisfy the corresponding (binary) clock \emph{relation}.
In other words, an n-ary clock \emph{relation} can be seen as the conjunction of $\frac{n(n-1)}{2}$ corresponding binary \emph{relations}.
For example, the ternary {\small\gt{causality}} \emph{relation} among $c_1$, $c_2$ and $c_3$ limits that
the three binary \emph{relations}, i.e., $c_1\textcolor{red}{\preceq} c_2$, $c_1\textcolor{red}{\preceq} c_3$ and $c_2\textcolor{red}{\preceq} c_3$, must be satisfied at the same time.

The probabilistic n-ary \emph{relation} is satisfied if the probability of the corresponding n-ary \emph{relation} being satisfied is greater than or equal to a given probability threshold $p$.

\begin{mydef}[{{Probabilistic N-ary Relation}}]
The probabilistic $n$-ary relation among $c_1$, $c_2$, ..., $c_n$, denoted as $\textcolor{red}{\sim_p}(c_1, c_2, ..., c_n)$, is satisfied if the following condition holds:
\[\mathcal{M} \vDash \textcolor{red}{\sim_p}(c_1, c_2, ..., c_n) \Longleftrightarrow Pr(\textcolor{red}{\sim}(c_1, c_2, ..., c_n))\geq p\]
\noindent where $\textcolor{red}{\sim} \in \{\textcolor{red}{\subseteq, \equiv, \prec, \preceq, \#} \}$, $Pr(\textcolor{red}{\sim}(c_1, c_2, ..., c_n))=\frac{1}{k} \sum\limits_{j=1}^{k} \left\{R_j \models \textcolor{red}{\sim_p}(c_1, c_2, ..., c_n) \right\}$, which is the probability of the $n$-ary \emph{relation} being satisfied, i.e., the ratio of {{runs}} satisfying the n-ary \emph{relation}
among the total number of $k$ {{runs}}.
\end{mydef}

\section{Translation of PrCCSL$^*$ into UPPAAL-SMC Models}
\label{sec:translation}

In Pr\ccsl$^*$, an automotive system can be specified as a {\small\gt{PCBS}} and the requirements of the system are specified as clock \emph{relations}.
To enable the formal analysis of system behaviors/requirements specified in Pr\ccsl$^*$,
in this section, we first present how to translate Pr\ccsl$^*$ elements, including {\small\gt{Dense}}{\small\gt{ClockType}}, \emph{expressions}, {\small\gt{(probabilistic) clock actions}} and \emph{relations}, into verifiable \uppaal\ models.
Then, we introduce our developed tool ProTL for supporting automatic translation and formal verification of Pr\ccsl$^*$ specifications.

\vspace{0.05in}

\noindent\textbf{Clock and DenseClockType}
In Pr\ccsl$^*$, a clock is either a logical clock or a dense clock. A logical clock ${c}$ represents an event, which is represented  as a

\label{sec:clockhis}
\begin{wrapfigure}{R}{2in}
  \vspace{-0.1in}
  \includegraphics[width=2in]{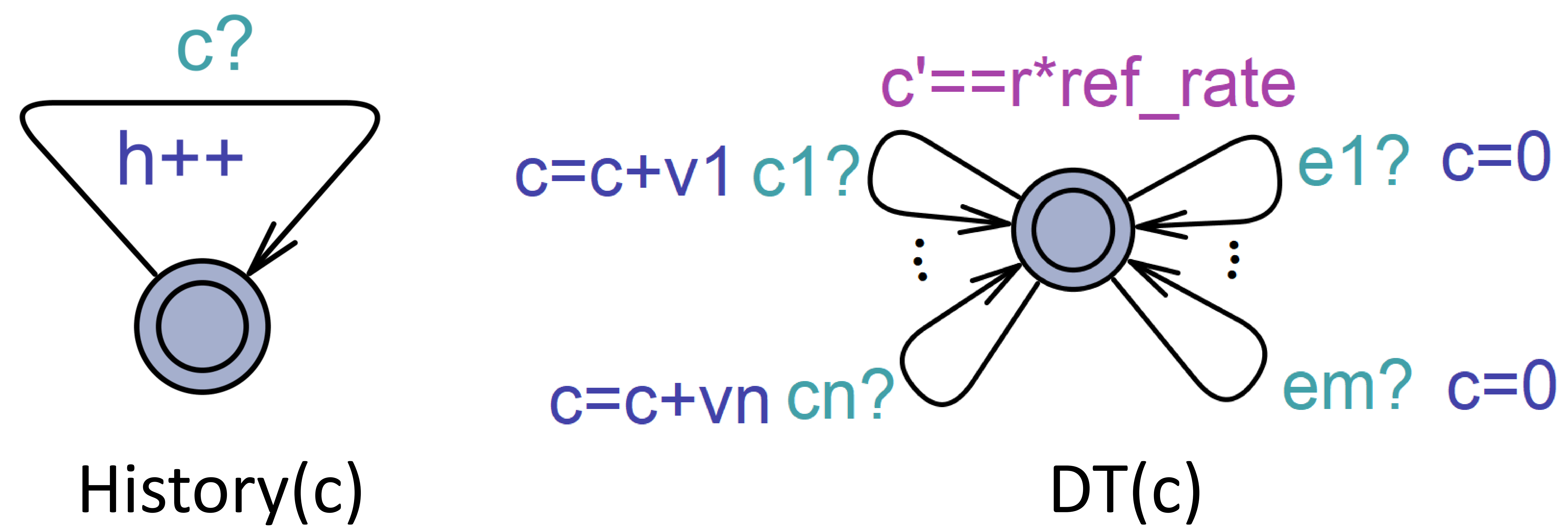}
  \caption{History and DenseClockType}
  \vspace{-0.1in}
 \label{fig:clock-type}
\end{wrapfigure}

\noindent \emph{synchronization channel} ${c!}$ in \smc.
The history of ${c}$ (which represents the number of times that $c$ has ticked currently) is modeled as the {\small\gt{History(c)}} STA (stochastic timed automata) in Fig. \ref{fig:clock-type}: whenever ${c}$ occurs (${c?}$), the value of its history is increased by 1 (i.e., $h$++).

A dense clock in Pr\ccsl$^*$ represents the physical time, which is considered as a continuous and
unbounded progression of time instants. A dense clock is represented as a ``\emph{clock}'' variable in \smc \cite{smc}, which is a real type variable increasing monotonically with a certain rate. For instance, the \emph{idealClk} is mapped to a standard \emph{clock} variable whose  increase rate is 1 in \smc.

To describe the continuous and instaneous variations of physical quantities (e.g., energy consumption, temperature), those quantities are represented as dense clocks instantiated by different {\small\gt{DenseClockTypes}}. A {\small\gt{DenseClockTypes}} $DT$ is defined based on the  {\small\gt{reference}}, {\small\gt{factor}}, {\small\gt{offset}} and {\small\gt{reset}} attributes (see Definition 3.4).
{\small\gt{DT(c)}} STA in Fig. \ref{fig:clock-type} is a generic representation of $DT$ in \smc, in which $c$ is a dense clock instantiated from $DT$.
The {\small\gt{factor}} of $DT$ (denoted $r$) is the rate of $c$ compared to the {\small\gt{reference}} clock $ref$. Let $ref\_rate$ represent the increase rate of $ref$ with respect to ideal physical time (i.e., $idealClk$). Thus the increase rate of $c$ compared to $idealClk$ equals to $r*ref\_rate$, modeled as the \emph{invariant} ``$c'$==$r*ref\_rate$'' in Fig. \ref{fig:clock-type}.
Thus, $c$ is changed continuously according to the differential equation ``$c$=$\int r*ref\_rate\ dt$'',  where $t\in[0,\ \infty]$ represents the physical time.
Moreover,  {\small\gt{offset}} specifies the instaneous changes of $c$ activated by a set of events $c_1, c_2, ..., c_n$. As shown in Fig. \ref{fig:clock-type}, each tuple in {\small\gt{offset}} corresponds to a self-loop transition where a discrete increment of $c$ is performed.
{\small\gt{reset}} is a set of events whose occurrences reset the time value of $c$ into 0, modeled as the transitions where $c$ is reset.

 \begin{wrapfigure}{R}{1.1in}
   \vspace{-0.1in}
  \includegraphics[width=1.1in]{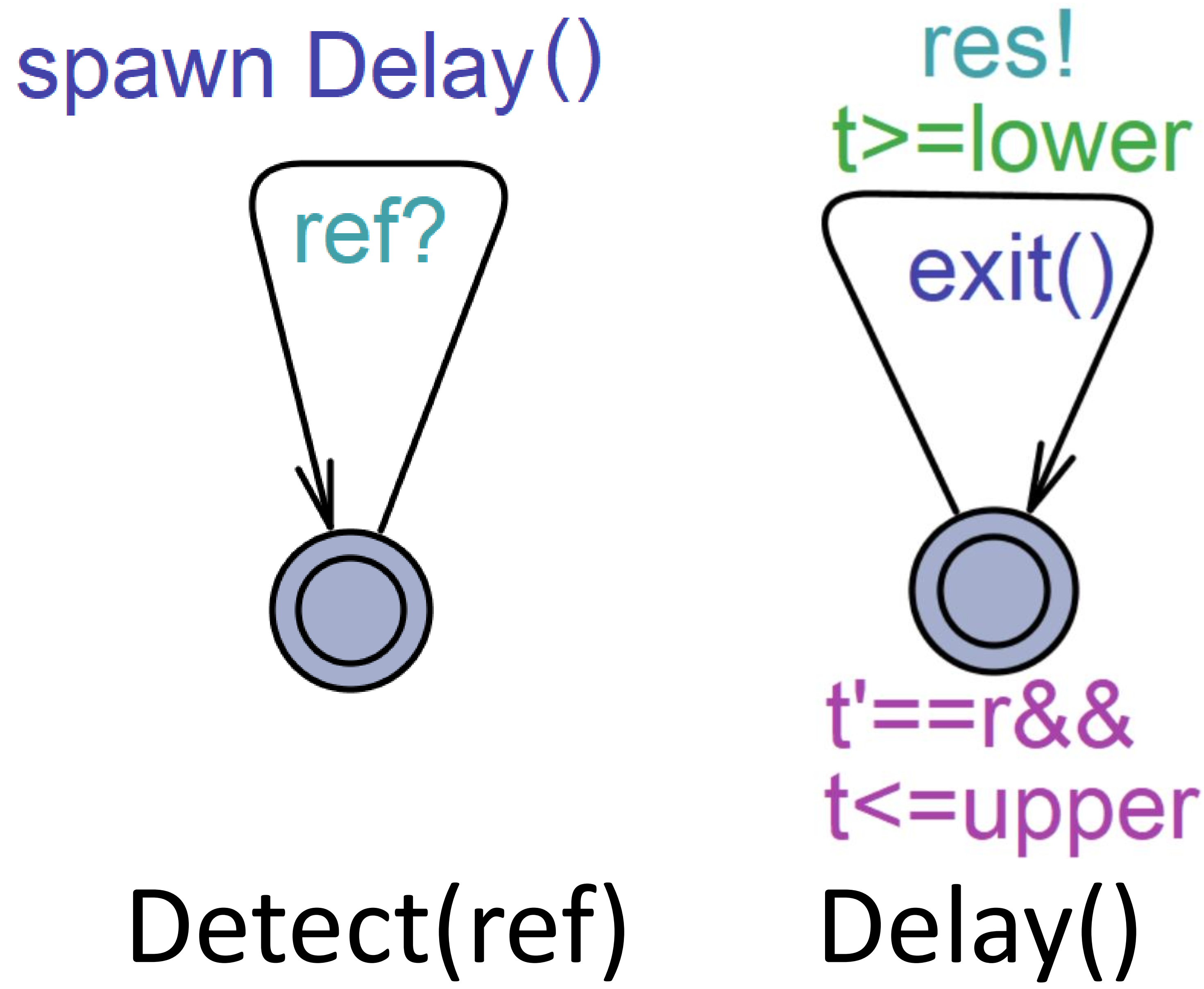}
  \caption{DelayFor}
    \vspace{-0.1in}
 \label{fig:delayFor}
 \end{wrapfigure}

\vspace{0.05in}
\noindent\textbf{Clock Expressions} generate new clocks based on existed clocks.  To model clock \emph{expressions} in \smc,  the resulting clock of the \emph{expression} is represented by a channel variable $res$ and the semantics of the \emph{expression} is modeled as an STA that determines when $res$ ticks (via $res!$).
For example, the interval-based {\small\gt{DelayFor}} \emph{expression} (Definition 3.1), expressed as ${res}$ $\triangleq$ $ref$ $([lower, upper])$ $\textcolor{red}{\rightsquigarrow}$ ${base}$, is modeled as the STA in Fig. \ref{fig:delayFor}. {\small \gt{DelayFor}}  defines a new clock $res$ based on a \emph{reference} clock ($ref$) and a $base$ clock, i.e., $\forall i \in N^+$, the $i^{th}$ tick of $res$ is generated by delaying the $i^{th}$ tick of $ref$ for $[lower,\ upper]$ units of $base$.
The generation process of the $i^{th}$ tick of $res$ can be summarized as: when the $i^{th}$ tick of $ref$ occurs,
after $[lower,\ upper]$ time units of $base$ clock is elapsed, the $i^{th}$ tick of $res$ is triggered.
The generation processes of different ticks of $res$ are independent and can run in parallel.
Therefore, we model the generation of each tick of $res$ as a spawnable STA \cite{david2015uppaal} (i.e., the {\small\gt{Delay()}} STA in Fig. \ref{fig:delayFor}), which is dynamically spawned (by {\small\gt{Detect(ref)}} STA) whenever $ref$ occurs ($ref?$), and terminated when the calculation of the current tick of $res$ is completed (denoted ``$exit()$'').
Here the $base$ in {\small \gt{DelayFor}} is a dense clock and $r$ is the increase rate of $base$ compared to $idealClk$.  For the STA of {\small \gt{DelayFor}} with discrete $base$ clock and other clock \emph{expressions}, refer to \cite{protl}.

\vspace{0.05in}

\begin{wrapfigure}{R}{2in}
   \vspace{-0.1in}
  \includegraphics[width=2in]{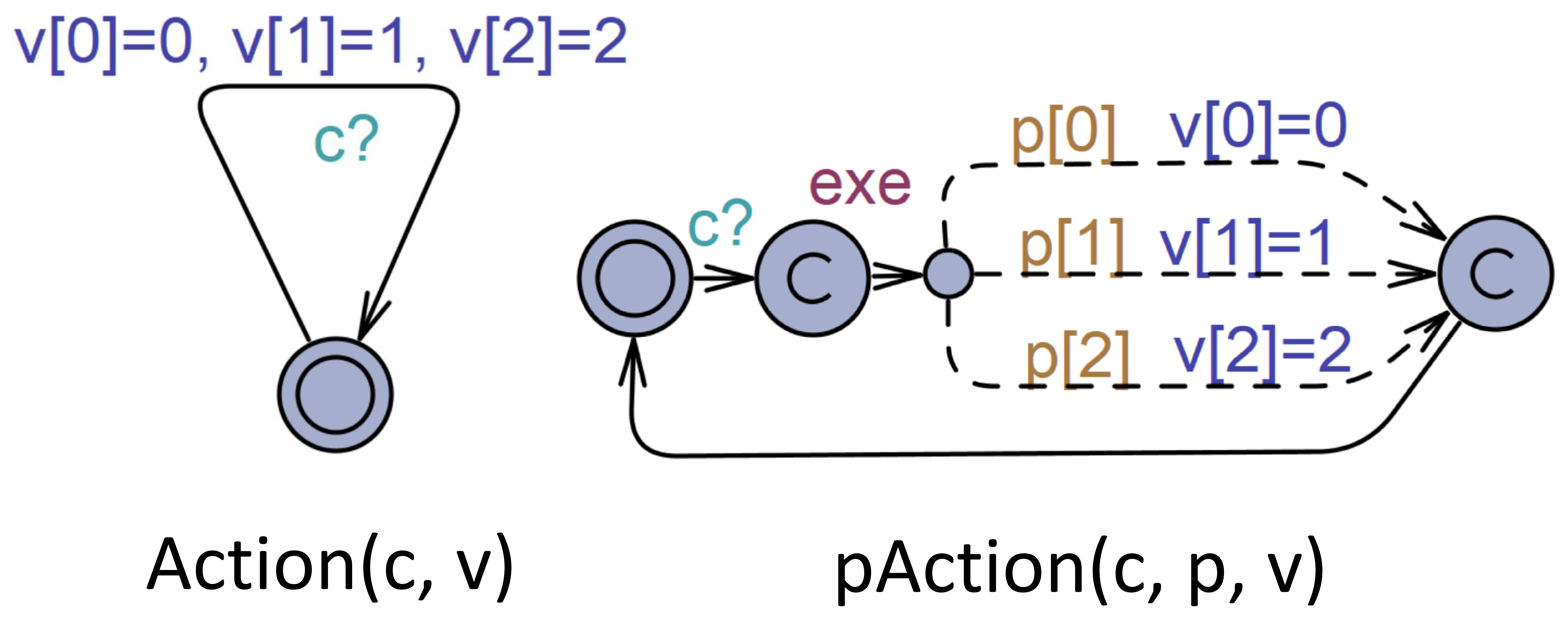}
   \caption{STA of {\small\gt{ clock action}}}
     \vspace{-0.1in}
 \label{fig:action}
 \end{wrapfigure}

\noindent\textbf{(Probabilistic) Clock Action} (see Definition 3.2 and 3.3) of clock $c$
are modeled as the STA in Fig. \ref{fig:action}.
The {\small\gt{clock action}} ``$c$ \textcolor{red}{$\rightarrow$} $ \{v[0]=0,\ v[1]=1,\ v[2]=2\}$'' is modeled as the $Action(c,\ v)$ STA: when $c$ ticks ($c?$), the assignment operations on the integer array $v$ are executed.
The {\small\gt{probabilistic clock action}} of clock $c$ specifies a set of actions that are performed when $c$ ticks based on a discrete probability distribution. For instance, the {\small\gt{probabilistic clock action}}
 \noindent``$c\ \textcolor{red}{\rightarrow_p} \{(p[0], v[0]=0), (p[1], v[1]=1), (p[3], v[2]=2)\}$'' is modeled as the $pAction(c,\ p,\ v)$ STA, in which each element of the {\small\gt{action}} is mapped to a probabilistic transition weighted by corresponding probability.

Based on the mapping patterns described above, a {\small\gt{PCBS}} (that consists of a set of clocks, {\small\gt{clock types}}, \emph{expressions} and {\small\gt{clock actions}}) can be represented as a network of STA (NSTA), which consists of the STA of corresponding {\small\gt{clock type}}, {\small\gt{(probabilistic) clock actions}} and \emph{expressions}.

\vspace{0.05in}
\noindent\textbf{Probabilistic Clock Relations}
To represent Pr\ccsl$^*$ \emph{relations} in \smc , observer STA that capture the semantics of standard
{\small\gt{subclock}}, {\small\gt{coincidence}}, {\small\gt{exclusion}}, {\small\gt{precedence}} and {\small\gt{causality}}
 \emph{relations} are constructed.
\begin{table*}[t]
\centering
\caption{STA of PrCCSL$^*$ Relations}
\begin{tabular}{cp{13.1cm}}
\toprule  %
STA  &Remarks  \\
\midrule  %
   \begin{minipage}{1.5in}
      \includegraphics[width=1.5in]{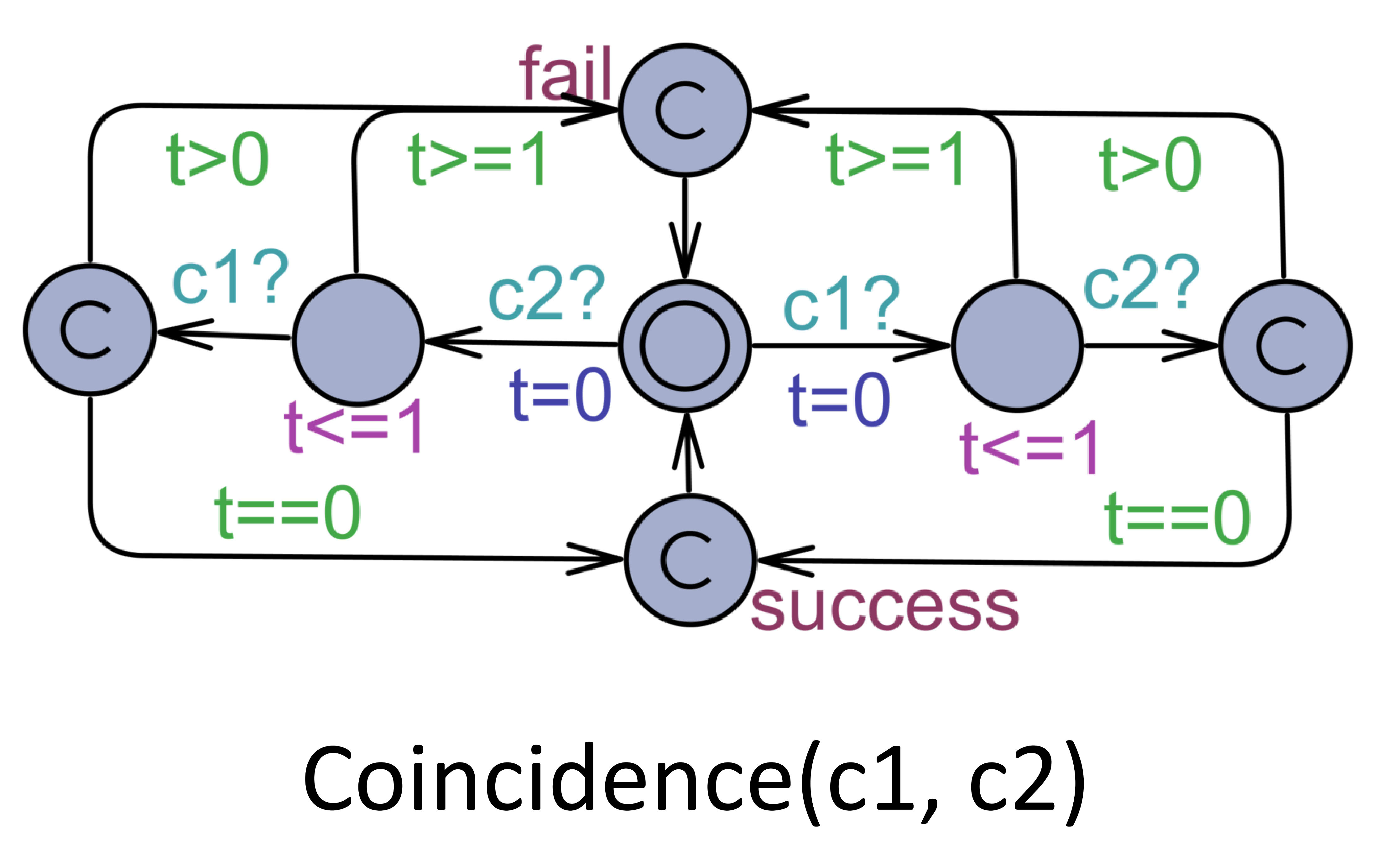}
    \end{minipage}

& \tabincell{l}{{Coincidence} \emph{relation} $c1$ \textcolor{red}{$\equiv_p$} $c2$ delimits that two clocks must tick simultaneously.
When $c1$ ($c2$) ticks via $c1?$ \\($c2?$), the STA
judges if the other clock, $c2$ ($c1$) ticks at the same time. If there is no time elapsed between \\the corresponding occurrences of $c1$ and $c2$ (i.e., ``t==0''), the STA transits to $success$ location. Otherwise, \\it goes to $fail$ location.
}
\\
\hline
   \begin{minipage}{1.5in}
      \includegraphics[width=1.5in]{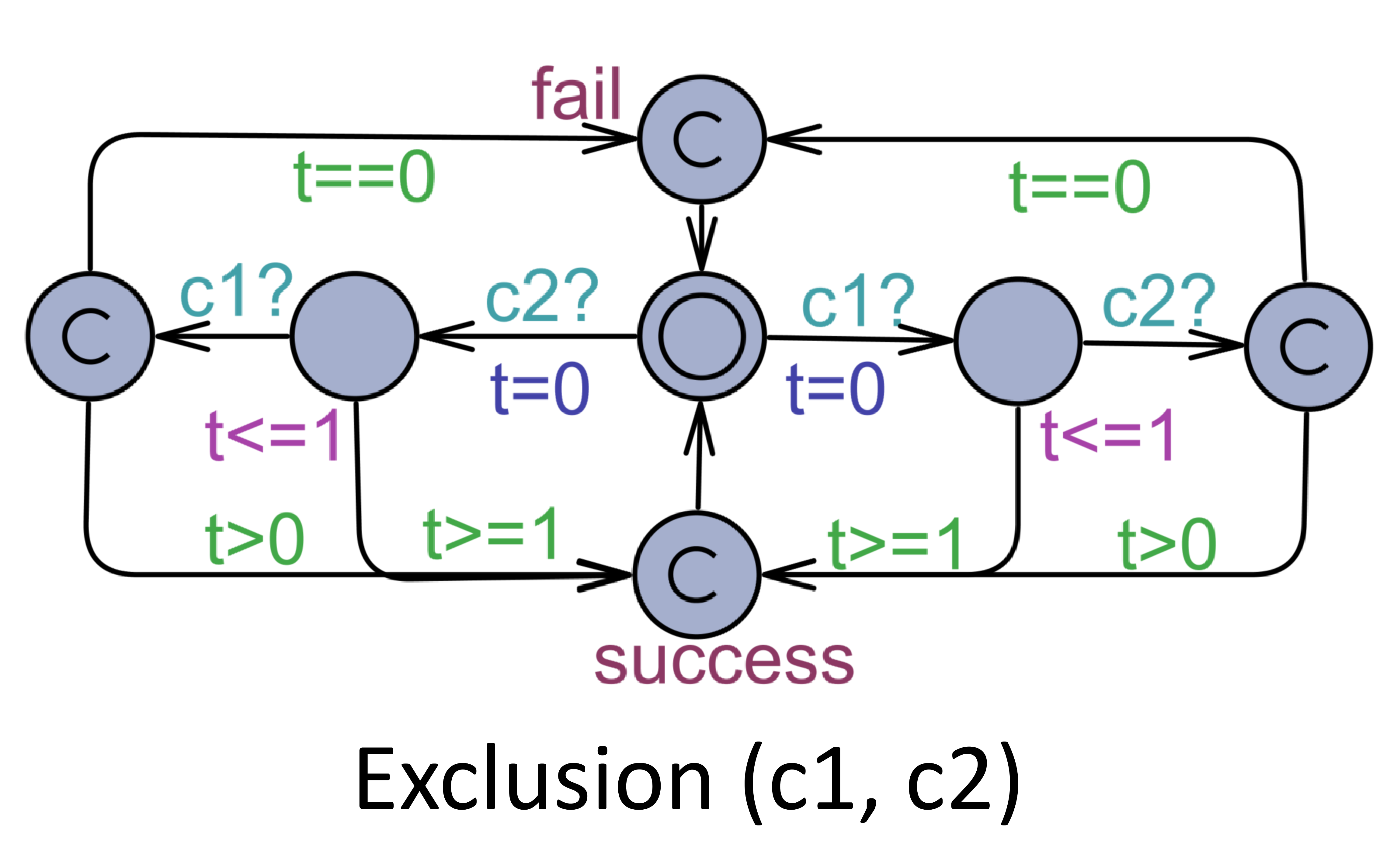}
    \end{minipage}

& \tabincell{l}{ Exclusion \emph{relation} $c1$ \textcolor{red}{$\#_p$} $c2$ limits that two clocks must not occur at the same time.
Contrary to the\\ ${Coincidence(c1,\ c2)}$ STA, when $c1$ ($c2$) ticks and if the other clock ticks simultaneously (``t==0''), the STA \\goes to the $fail$ location.}
\\
\hline
   \begin{minipage}{1.5in}
      \includegraphics[width=1.5in]{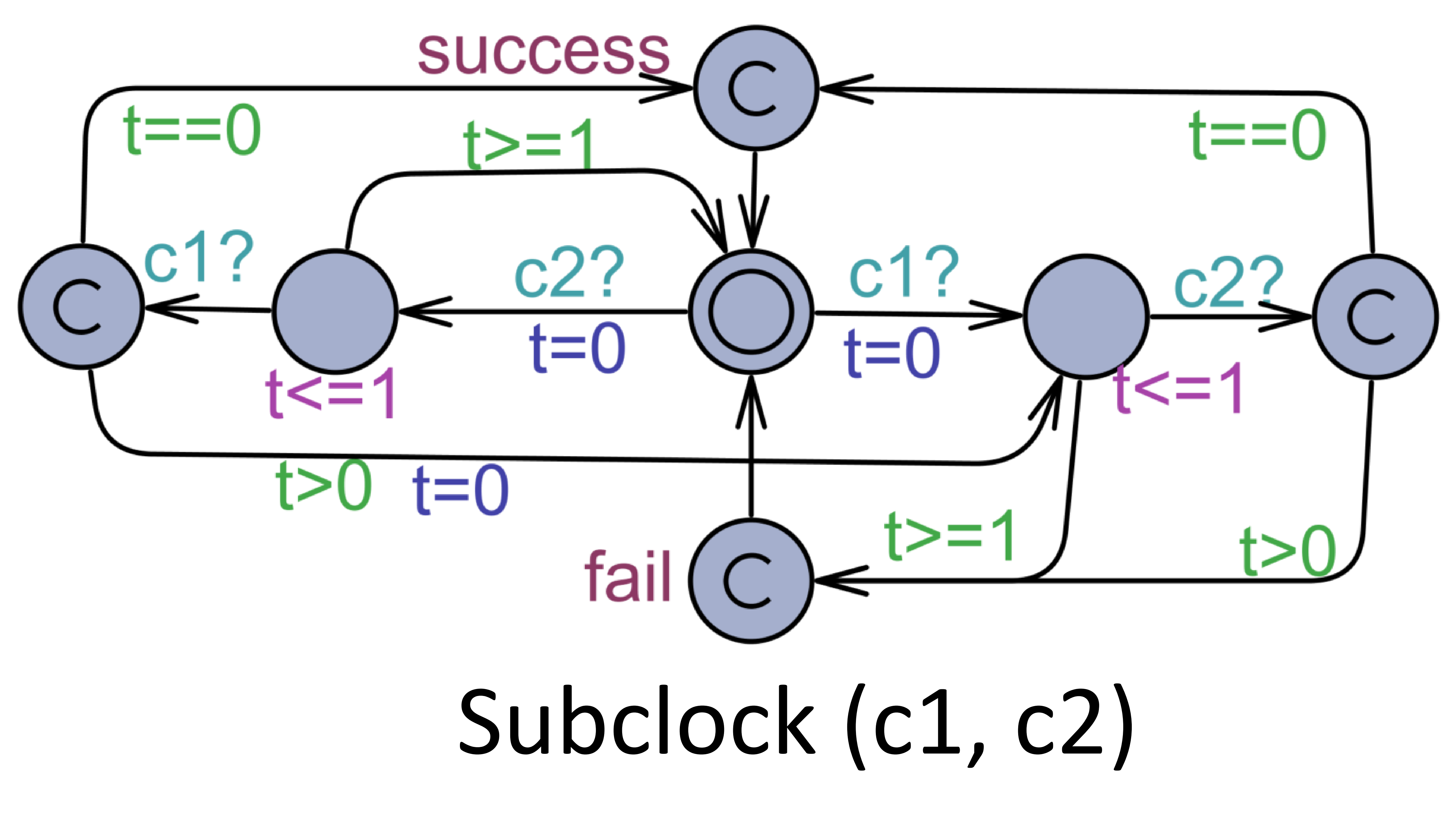}
    \end{minipage}

& \tabincell{l}{{Subclock} \emph{relation} $c1$ \textcolor{red}{$\subseteq_p$} $c2$  states that $c2$ (superclock) must tick when
 $c1$ (subclock) ticks, which is interpreted as\\ a conditional  {{coincidence}} \emph{relation}: when $c1$ ticks, $c2$ must coincides with $c1$.
Similar to $Coincidence(c1,\ c2)$ \\STA, when $c1$ ($c2$) occurs,
the $Subclock(c1,\ c2)$ STA checks whether the other clock also ticks at the same \\instant.
If $c1$ ticks and $c2$ does not occur (denoted ``t$>$0''),
the relation is violated and the STA transits to $fail$\\ location.

}
\\
\hline
  \begin{minipage}{1.5in}
      \includegraphics[width=1.5in]{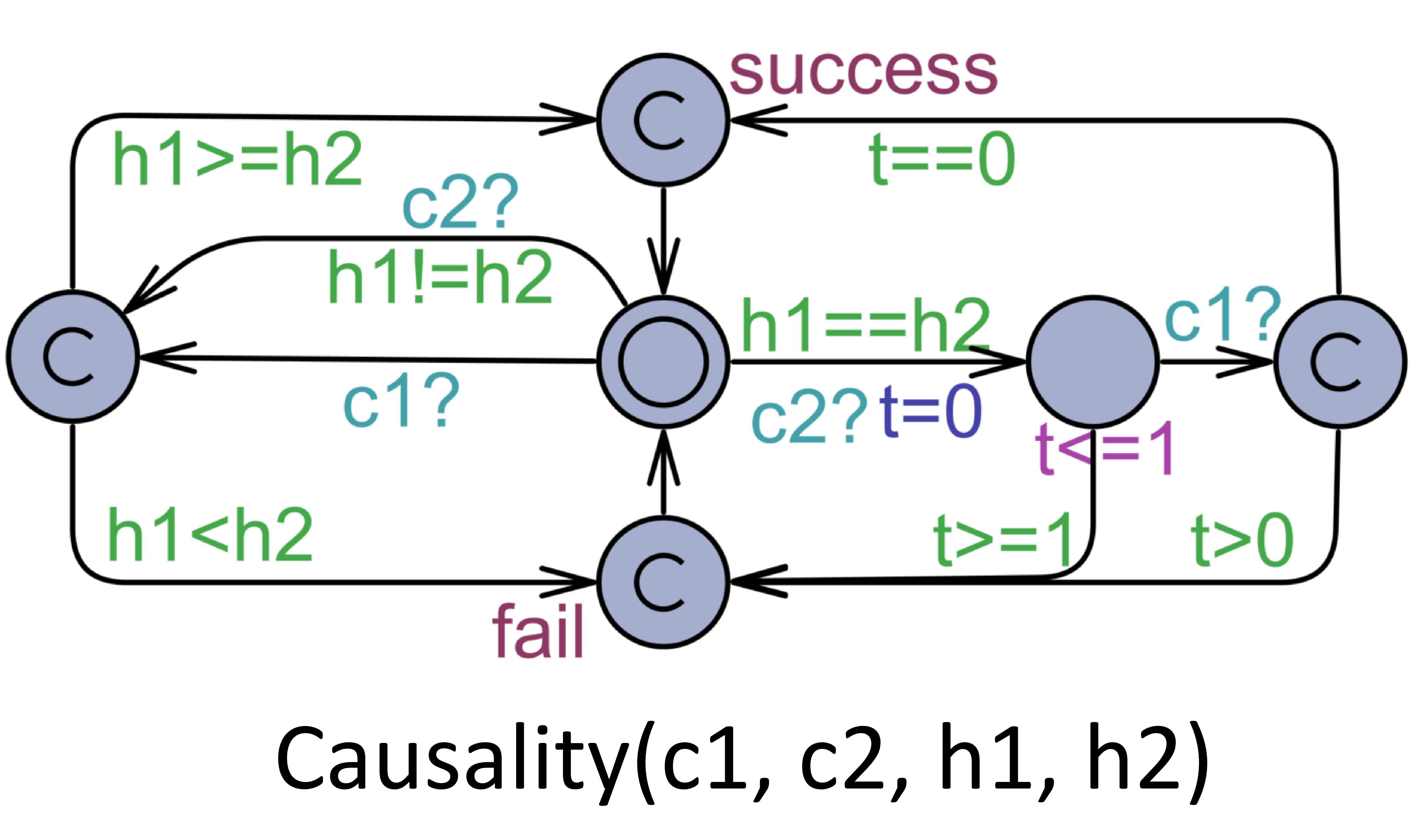}
    \end{minipage}

& \tabincell{l}{{Causality} \emph{relation} $c1$ \textcolor{red}{$\preceq_p$} $c2$ states that $c2$ (\emph{effect})
must not tick prior to $c1$ (\emph{cause}). When $c1$ or $c2$ ticks, if the\\ two clocks are coincident (represented by ``t==0'') or $c1$ ticks faster (denoted ``$h1\geq h2$''), the \emph{relation} is satisfied \\and the STA goes to $success$ location. Otherwise, the STA goes to $fail$ location.

}
\\
\hline
   \begin{minipage}{1.5in}
      \includegraphics[width=1.5in]{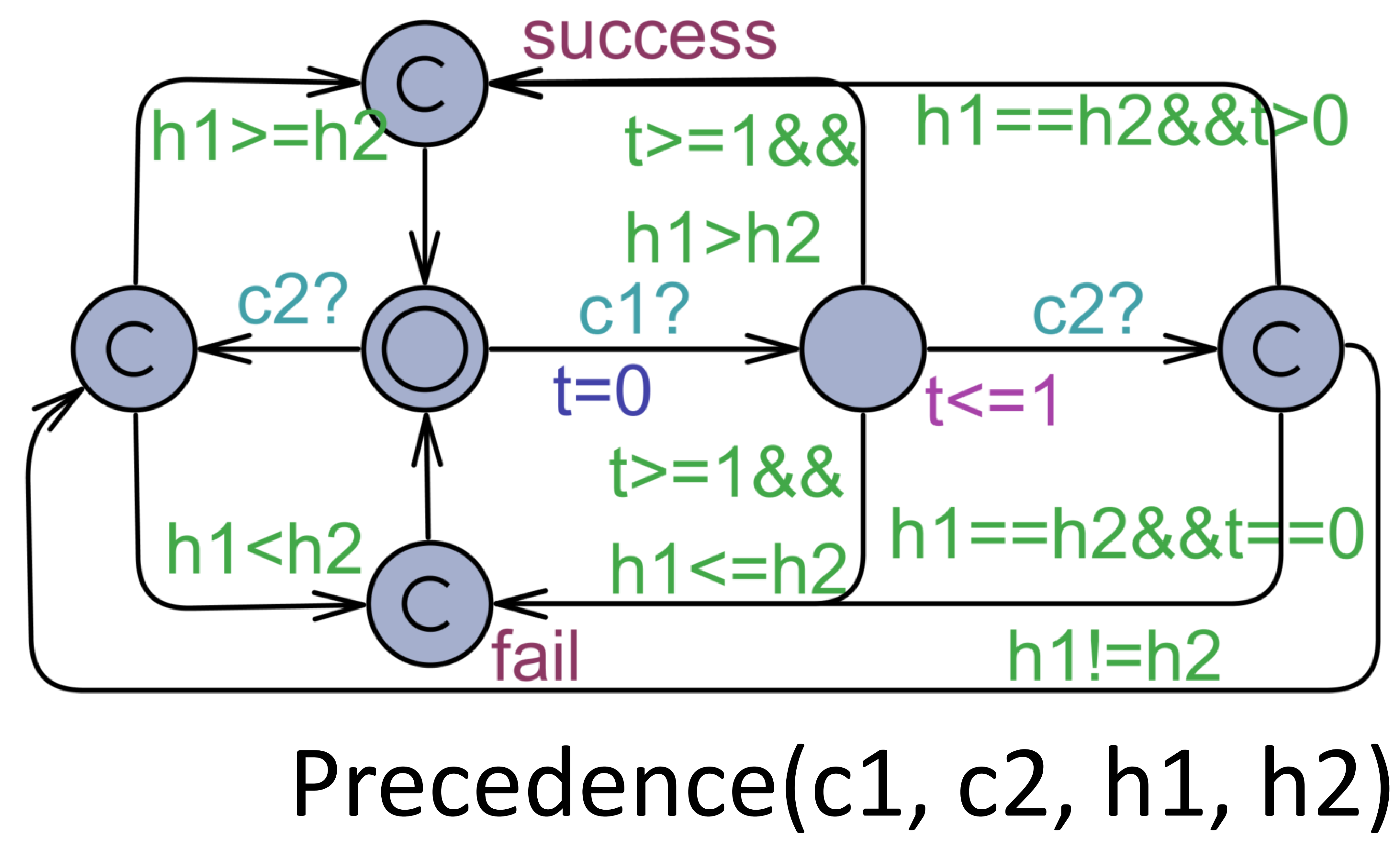}
    \end{minipage}

& \tabincell{l}{{Precedence} \emph{relation} $c1$ \textcolor{red}{$\prec_p$} $c2$ states that $c1$ must run faster than $c2$.
When $c1$ or $c2$ ticks, if $c2$ ticks faster\\ (``h1$<$h2'') or $c1$ and $c2$ are coincident (represented by ``h1==h2\&\&t==0'')
the \emph{relation} is violated and $fail$\\ location is activated.
}
\\

\bottomrule %
\end{tabular}
  \label{table:relation-sta}%
\end{table*}
 In our earlier work \cite{ifm18}, we have shown the translation patterns from clock \emph{relations} into STA. However, the patterns are given based on discrete time semantics, i.e., the continuous physical time line is discretized into a set of equalized steps.
As a result, (in the extreme cases) two clock instants are still considered coincident even if they are one time step apart.
In this paper, we refine the STA of \emph{relations} in \cite{ifm18} to support the translation of \emph{relations} conformed to continuous time semantics.
The refined STA are illustrated in Table \ref{table:relation-sta}, in which $t$ represents the time delay between
  two consecutive instants of clock $c1$ and $c2$. $c1$ and $c2$ are simultaneous if $t$ equals to 0.  $h1$ and $h2$ represent the histories of $c1$ and $c2$, respectively. Each \emph{relation} is mapped to an observer STA that contains a ``$fail$'' location, which suggests the violation of corresponding \emph{relation}.
Recall the definition of probabilistic \emph{relations} in Sec. \ref{sec:preliminary},
the probability of a \emph{relation} being satisfied is interpreted as a ratio
of runs that satisfies the \emph{relation} among all runs.
It is specified as \emph{Hypothesis Testing} query in \smc,
$H_0$:  $\frac{m}{k} \geq p$ against $H_1$: $\frac{m}{k} < p$, where $m$ is the number of runs
satisfying the given \emph{relation} out of all $k$ runs. As a result, the probabilistic \emph{relations}
 are interpreted as the query: $Pr[bound]([\ ]\ \neg STA_{obs}.fail)\geq p$, which means that the probability of
  the ``$fail$'' location of the observer STA (denoted $STA_{obs}$) never being reached should be greater than or equal to threshold $p$.

\begin{figure*}[htbp]
  \centering
  \includegraphics[width=6.2in]{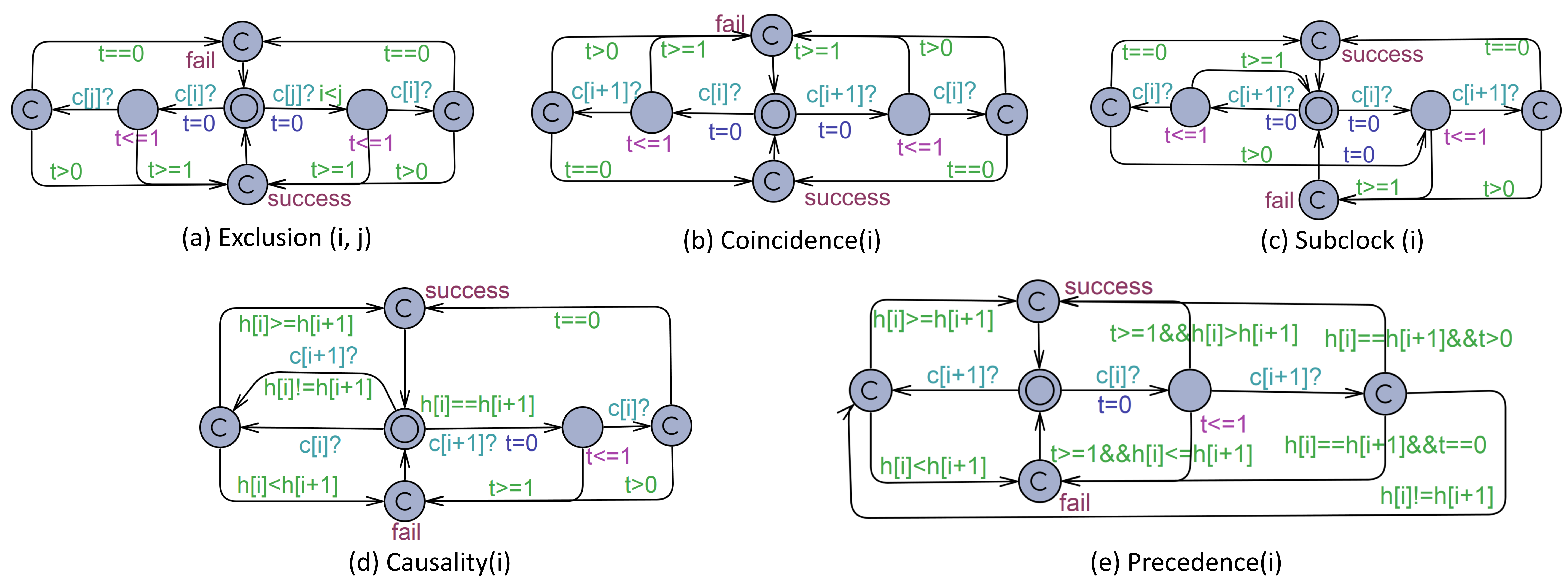}
  \caption{STA of N-ary Relations}
  \label{fig:nary-relation}
\end{figure*}

Requirements associated with multiple clocks can be expressed by n-ary \emph{relations} (see Definition 3.6).
According to the definition, an n-ary \emph{relation} among $n$ clocks $c_1$, $c_2$, ..., $c_n$ is satisfied if
 any pair of two clocks $\langle c_i$, $c_j \rangle$ ($1\leq i<j\leq n$) satisfies the corresponding binary \emph{relation}.
Since an n-ary clock \emph{relation} can be seen as the conjunction of $\frac{n(n-1)}{2}$ binary \emph{relations}, we construct an STA for the \emph{relation} of each pair $\langle c_i$, $c_j\rangle$ ($1\leq i<j\leq n$) and an n-ary \emph{relation} is represented as the synchronization product of the $\frac{n(n-1)}{2}$ STA.
For instance, an n-ary {\small\gt{exclusion}} can be represented as the composition of the $\frac{n(n-1)}{2}$ {\small\gt{Exclusion(i, j)}} STA in Fig. \ref{fig:nary-relation}(a) (similar to the {\small\gt{Exclusion(c1, c2)}} STA in Table \ref{table:relation-sta}), which represents the {\small\gt{exclusion}} \emph{relation} between $c_i$ and $c_{j}$.
The probabilistic n-ary {\small\gt{exclusion}} is specified as:
$Pr[\leq bound] ([\ ] forall (i:int[1, n])\ forall (j:int[1, n])\ (i<j\ imply\ not\ Exclusion(i, j).fail))\geq p$.

The {\small\gt{coincidence}}, {\small\gt{subclock}},  {\small\gt{precedence}} and {\small\gt{causality}} \emph{relations} are transitive \emph{relations}, i.e., if both the \emph{relations} between $c_i$ and $c_j$ and the \emph{relation} between clock $c_j$ and $c_k$ are satisfied, then the \emph{relation} between $c_i$ and $c_k$ is also satisfied. Thus, an n-ary transitive \emph{relation} can be interpreted as the combination of $n-1$ binary \emph{relations}.
For instance, the n-ary  {\small\gt{coincidence}} \emph{relation} is interpreted as: $\textcolor{red}{\equiv}(c_1,\ c_2,\ ...,\ c_n)$ $\Longleftrightarrow \bigwedge\limits_{i=1}^{n-1} c_i\textcolor{red}{\equiv}c_{i+1}$.
As illustrated in Fig. \ref{fig:nary-relation}(b), {\small\gt{Coincidence(i)}} STA represents the {\small\gt{coincidence}} \emph{relation} between $c_i$ and $c_{i+1}$ (similar to the {\small\gt{Coincidence(c1, c2)}} STA in Table \ref{table:relation-sta}). The n-ary {\small\gt{coincidence}} \emph{relation} can be represented as the composition of the $n-1$ {\small\gt{Coincidence(i)}} STA, where $i\in\{1,\ 2,\ 3,\ ..., n-1\}$. The n-ary probabilistic {\small\gt{coincidence}} \emph{relation} can be verified by using the query: $Pr[\leq bound] ([\ ] forall (i:int[1, n-1]) (not\ Coincidence(i).fail))\geq p$.
Similarly, the other three transitive \emph{relations}, i.e., {\small\gt{subclock}},  {\small\gt{precedence}} and {\small\gt{causality}} \emph{relations}, can be represented based on the STA in Fig. \ref{fig:nary-relation}.

\vspace{0.05in}

\noindent\textbf{Tool Support}
To improve the efficiency and accuracy of translation, we implement a tool \protl\ (Probabilistic \ccsl\ TransLator) \cite{protl}, which
provides a push-button transformation from Pr\ccsl$^*$ specifications into \smc\ models.
Moreover, to enable the formal verification of Pr\ccsl$^*$ specifications,
\protl\ brings the capability of verification for the translated \smc\ models by employing \smc\ as its verification backend.
Furthermore, \protl\ offers a configuration panel for customizable generation of five types of probabilistic queries (introduced in Sec. \ref{sec:preliminary}). \protl\ can also generate counter-examples that depict the evolution of clocks related to the unsatisfied clock \emph{relations}, which provide diagnosis information for further refinement of Pr\ccsl$^*$ specifications.

\section{Case Studies}
\label{sec:case-study}
We show the applicability of \protl\ on two automotive systems case studies.
We report only the verification on a list of representative requirements for each example and further details can be found in \cite{protl}.

\vspace{0.05in}
\noindent \textbf{Autonomous Vehicle} (AV) \cite{ifm18} reads the road signs, e.g., ``speed limits'', ``stop'' or ``right/left turn'', and adjusts its speed and movements accordingly.  To achieve traffic sign recognition, a camera is equipped in the vehicle to
capture images. The camera relays the captured images to a sign recognition device periodically.
The representative requirements on AV are listed below:

\noindent A1. A periodic acquisition of camera must be carried out every 50ms with a jitter of 10ms.

\noindent A2. If the vehicle detects a left turn sign, it should start to turn left within 300ms.

\noindent A3. The detected image should be computed within [20, 30]ms in order to generate the desired sign type.

\noindent A4. When a traffic sign is recognized, the speed of vehicle should be updated within [50, 150]ms based on the sign type.

\noindent A5. The required environmental information should arrive to the controller within 40 ms, i.e., the input signals (traffic sign type, speed,  direction, gear and torque) must be detected by controller within 40 ms.

\noindent A6. The execution time interval from the controller to the actuator must be less than or equal to the sum of the worst case execution time intervals of controller and actuator.

\noindent A7. While the vehicle turns left, the ``turning right'' mode should not be activated. The events of turning left and right are considered as exclusive and specified as an exclusion constraint.

We specify the system behaviors of AV as a {\small\gt{PCBS}} and requirements as clock \emph{relations} in Pr\ccsl$^*$.
The Pr\ccsl$^*$ specifications of requirements (A1--A7) and the verification results are illustrated in Table \ref{table:av}.
Let $camera$ represent the triggering event of camera and $camera[i]$ denote the $i^{th}$ occurrence of $camera$.
A1 can be interpreted as: $\forall i\ \in \mathbb{N}^{+}$, $camera[i+1]$ should
occur later than $camera[i]$ delaying for 40ms but prior to $camera[i]$ delaying for 60ms, which is specified as a ternary {\small\gt{precedence}} \emph{relation} in Table \ref{table:av}. The ``forall'' query in \uppaal\ is employed to verify the ternary \emph{relation}.

\begin{table*}[t]
  \scriptsize
  \centering
   \renewcommand\arraystretch{1.5}
  \caption{Verification Results of AV}
    \begin{tabular}{|c|c|p{220pt}|c|c|c|c|c|}
    \hline
    Req & Spec & Expression & Runs & Result & Time (Min) & Mem (Mb) & CPU (\%)   \\
    \hline
    \multirow{2}{*}{A1}   &Pr\ccsl$^*$& \tabincell{l}{{$cameraFltr \triangleq camera\ \textcolor{red}{\blacktriangledown} 01(1)$}\\
{$cameraDelay40 \triangleq camera\ (40)\ \textcolor{red}{\rightsquigarrow}\ ms$}\\
{$cameraDelay60 \triangleq camera\ (60)\ \textcolor{red}{\rightsquigarrow}\ ms$}\\
{$\textcolor{red}{\prec_{0.96}}\ (cameraDelay40, cameraFltr, cameraDelay60)$}}  &142 & valid & 9.69 & 8.49& 25.08\\ \cline{2-3}
    & UPPAAL &  \tabincell{l}{Pr[$\leqslant$10000]([ ] forall i:[1, 2] $\neg A1\_Precedence(i).fail$) $\geq 0.96$} & & & & & \\
   \hline
    \multirow{2}{*}{A2}   &Pr\ccsl$^*$& \tabincell{l}{{$leftSignDe\ \triangleq\ detectLeftSign\ (300)\ \textcolor{red}{\rightsquigarrow}\ ms$}\\ {$startTurnLeft\ \textcolor{red}{\prec_{0.95}}\ leftSignDe$}}  & 161 & valid & 8.72 & 8.30 & 25.40 \\\cline{2-3}
    & UPPAAL & \tabincell{l}{Pr[$\leqslant$10000]([ ] $\neg A2\_Precedence.fail$) $\geq 0.95$} & & & & &\\
   \hline
   \multirow{2}{*}{A3}   &Pr\ccsl$^*$& \tabincell{l}{{$ImgRecDe20\ \triangleq\ ImgRec\ (20)\ \textcolor{red}{\rightsquigarrow}\ ms$}\\
   $ImgRecDe30\ \triangleq\ ImgRec\ (30)\ \textcolor{red}{\rightsquigarrow}\ ms$\\
   {$\textcolor{red}{\preceq_{0.96}}(ImgRecDe20,\ signType,\ ImgRecDe30)$}}  & 142 & valid & 10.31 & 8.51 & 25.27 \\\cline{2-3}
    & UPPAAL & \tabincell{l}{Pr[$\leqslant$10000]([ ] forall i:[1, 2] $\neg A3\_Causality(i).fail$) $\geq 0.96$} & & & & &\\
   \hline
   \multirow{2}{*}{A4}   &Pr\ccsl$^*$& \tabincell{l}{{$signTypeDe50\ \triangleq\ signType\ (50)\ \textcolor{red}{\rightsquigarrow}\ ms$}\\
   $signTypeDe150\ \triangleq\ signType\ (150)\ \textcolor{red}{\rightsquigarrow}\ ms$\\
   {$\textcolor{red}{\prec_{0.95}}(signTypeDe50,\ updateSpeed,\ signTypeDe150)$}}  & 140 & valid & 10.25 & 8.58 & 25.13 \\\cline{2-3}
    & UPPAAL & \tabincell{l}{Pr[$\leqslant$10000]([ ] forall i:[1, 2] $\neg A4\_Precedence(i).fail$) $\geq 0.95$} & & & & &\\
   \hline
   \multirow{2}{*}{A5}   &Pr\ccsl$^*$& \tabincell{l}{{$InfIn\ \triangleq\ speedIn\ \textcolor{red}{\wedge}\ posIn\ \textcolor{red}{\wedge}\ dirIn \textcolor{red}{\wedge}\  signType$}\\
   $SupIn\ \triangleq\ speedIn\ \textcolor{red}{\vee}\ posIn \textcolor{red}{\vee}\ dirIn\ \textcolor{red}{\vee}\ signType $\\
   $InfInDe40\ \triangleq\ InfIn\ (40)\ \textcolor{red}{\rightsquigarrow}\ ms$\\
   {$ SupIn\ \textcolor{red}{\preceq_{0.95}}\ InfInDe40$}}  & 140 & valid & 12.96 & 8.6 & 24.86 \\\cline{2-3}
    & UPPAAL & \tabincell{l}{Pr[$\leqslant$10000]([ ] $\neg A5\_Causality.fail$) $\geq 0.95$} & & & & &\\
   \hline
   \multirow{2}{*}{A6}   &Pr\ccsl$^*$& \tabincell{l}{{$signTypeDe\ \triangleq\ signType\ (SUM_{WCET})\ \textcolor{red}{\rightsquigarrow}\ ms$}\\
   {$actOut\ \textcolor{red}{\prec_{0.95}}\ signTypeDe$}}  & 140 & valid & 12.11 & 8.48 & 24.24 \\\cline{2-3}
    & UPPAAL & \tabincell{l}{Pr[$\leqslant$10000]([ ] $\neg A6\_Precedence.fail$) $\geq 0.95$} & & & & &\\
   \hline
   \multirow{2}{*}{A7}   &Pr\ccsl$^*$& \tabincell{l}{{$turnLeft \triangleq\ \{inLeft=1\}\ \textcolor{red}{?}\ always\ \textcolor{red}{:}\ never$}\\ {$turnRight\ \triangleq\ \{inRight\}\ \textcolor{red}{?}\ always\ \textcolor{red}{:}\ never$}\\
   {$turnLeft\ \textcolor{red}{\#_{0.95}}\ turnRight$}}  & 140 & valid & 8.82 & 8.40 &25.31 \\\cline{2-3}
    & UPPAAL & \tabincell{l}{Pr[$\leqslant$10000]([ ] $\neg A7\_Exclusion.fail$) $\geq 0.95$} & & & & &\\
   \hline

    \end{tabular}%
  \label{table:av}%
\end{table*}%
In the specification of A2, $detectLeftSign$ represents the event that a left turn sign is detected. $startTurnLeft$ denotes the event that the vehicle starts to turn left. We construct a new clock $leftSignDe$ by delaying $detectLeftSign$ for 300ms. A2 can be expressed as
 a {\small\gt{precedence}} \emph{relation} between $startTurnLeft$ and $leftSignDe$, i.e., $startTurnLeft$ should occur no later than
the occurrence of $leftSignDe$. Similarly, A3--A6 can be specified.
In the Pr\ccsl$^*$ specification of A7, $always$ ($never$) represents a clock that always (never) ticks.
Based on $always$ and $never$, we generate two new clocks  $turnLeft$ and $turnRight$ by using {\small\gt{ITE}} \emph{expressions}.
$turnLeft$ ($turnRight$) represents the event that the vehicle is turning left (right). A7 is specified as an {\small\gt{exclusion}} \emph{relation} between $turnLeft$ and $turnRight$.

\noindent \textbf{Cooperative Automotive System (CAS)} \cite{sac18} includes distributed and coordinated sensing, control, and actuation over three vehicles which are running in the same lane. A lead vehicle can run automatically by recognizing traffic signs on the road. The follow vehicle must set its desired velocity identical to that of its immediate preceding vehicle. Vehicles should maintain sufficient braking distance to avoid rear-end collision while remaining close enough to guarantee communication quality.
The position of each vehicle is represented by Cartesian coordinate $(x_i, y_i)$,
 where $x_i$ and $y_i$ ($i\in \{0, 1, 2\}$) are distances measured from the vehicle to the two fixed perpendicular lines,
 i.e., x-axis and y-axis, respectively.
A list of the representative requirements on CAS are given below:

\noindent B1. The follow vehicle should not overtake the lead vehicle.

\noindent B2. When the lead vehicle adjusts its movement (e.g., braking) regarding environmental information, the follow vehicle should move towards the lead one within 500ms.

\noindent B3. Each vehicle should maintain braking distance, i.e., if the braking distance is insufficient, the follow vehicle should decelerate within a given time.

\noindent B4. When the lead vehicle starts to turn left, both the lead and follow vehicle should complete turning and run in the same lane within a certain time.

\noindent B5. The velocity of vehicles should update every 30ms, i.e., a periodic acquisition of a speed sensor must be carried out for every 30ms.

\noindent B6. The required input signals of the environmental information (speed, position, direction) must be detected by controller within a given time window, i.e., 60ms.

\noindent B7. The controller of the follow vehicle should finish its execution within [30, 100]ms.

\begin{table*}[t]
  \scriptsize
  \centering
  \renewcommand\arraystretch{1.5}
  \caption{Verification Results of CAS}
    \begin{tabular}{|c|c|p{240pt}|c|c|c|c|c|}
    \hline
    Req & Spec & Expression & Runs & Result & Time (Min) & Mem (Mb) & CPU (\%)   \\
    \hline
    \multirow{2}{*}{B1}   &Pr\ccsl$^*$ &\tabincell{l}{{$runAtXDir \triangleq\ \{direction=xDir\}\ \textcolor{red}{?}\ always\ \textcolor{red}{:}\ never$}\\ {$overTake\ \triangleq\ \{x_1\geq x_0\}\ \textcolor{red}{?}\ always\ \textcolor{red}{:}\ never$}\\
{$runAtXDir\ \textcolor{red}{\#_{0.95}}\ overTake$}} & 140 & valid & 128.34 & 10.62 &25.34 \\ \cline{2-3}
    & UPPAAL &  \tabincell{l}{Pr[$\leqslant$10000]([ ] $\neg B1\_Exclusion.fail$) $\geq 0.95$} & & & & &\\
   \hline
    \multirow{2}{*}{B2}   &Pr\ccsl$^*$& \tabincell{l}{{$brakeDelay500\ \triangleq\ leadBrake\ (500)\ \textcolor{red}{\rightsquigarrow}\ ms$}\\
{$followBrake\ \textcolor{red}{\prec_{0.95}}\ brakeDelay500$}} &140 & valid &132.83 &8.15 &25.17 \\\cline{2-3}
    & UPPAAL & \tabincell{l}{Pr[$\leqslant$10000]([ ] $\neg B2\_Precedence.fail$) $\geq 0.95$} & & & & &\\
   \hline
   \multirow{2}{*}{B3}   &Pr\ccsl$^*$& \tabincell{l}{{$notSafe\ \triangleq\ \{inConst=true\&\&dist<safeDis\}\ \textcolor{red}{?}\ always\ \textcolor{red}{:}\ never$}\\
   $notSafeDe300\ \triangleq\ notSafe\ (300)\ \textcolor{red}{\rightsquigarrow}\ ms$\\
{$const2dec\ \textcolor{red}{\prec_{0.95}}\ notSafeDe300$}} & 140& valid & 126.40 &10.52 &25.29 \\\cline{2-3}
    & UPPAAL & \tabincell{l}{Pr[$\leqslant$10000]([ ] $\neg B3\_Precedence.fail$) $\geq 0.95$} & & & & &\\
   \hline
   \multirow{2}{*}{B4}   &Pr\ccsl$^*$& \tabincell{l}{
   $leadTurnLeftDe\ \triangleq\ leadTurnLeft\ (500)\ \textcolor{red}{\rightsquigarrow}\ ms$\\
  {$followTurn\ \textcolor{red}{\prec_{0.95}}\ leadTurnLeftDe$}}
   &54 & Unsatisfied& 60.87 &10.66 & 24.86 \\ \cline{2-3}
    & UPPAAL & \tabincell{l}{Pr[$\leqslant$10000]([ ] $\neg B4\_Precedence.fail$) $\geq 0.95$} & & & & &\\
   \hline
   \multirow{2}{*}{B5}   &Pr\ccsl$^*$& \tabincell{l}{{$prdClk \triangleq\ ms\ \textcolor{red}{\propto}\ 30$}\\
{$\textcolor{red}{\equiv_{0.98}}(leadSpeedTrig, prdClk, followSpeedTrig$)}} &145 & valid & 160.91 &10.69 & 25.05\\\cline{2-3}
    & UPPAAL & \tabincell{l}{Pr[$\leqslant$10000]([ ] forall (i:int[1,2]) $\neg B5\_Coincidence(i).fail$) $\geq 0.98$} & & & & &\\
   \hline
      \multirow{2}{*}{B6}   &Pr\ccsl$^*$& \tabincell{l}{{$InfIn\ \triangleq\ speedIn \textcolor{red}{\wedge}\ posIn \textcolor{red}{\wedge}\ dirIn $}\\
      {$SupIn\ \triangleq\ speedIn \textcolor{red}{\vee}\ posIn \textcolor{red}{\vee}\ dirIn $}\\
   $InfInDe\ \triangleq\ InfIn\ (60)\ \textcolor{red}{\rightsquigarrow}\ ms$
{$SupIn\ \textcolor{red}{\prec_{0.95}}\ InfInDe$}} &140 & valid & 188.63 &8.92 &24.09 \\\cline{2-3}
    & UPPAAL & \tabincell{l}{Pr[$\leqslant$10000]([ ] $\neg B6\_Precedence.fail$) $\geq 0.95$} & & & & &\\
   \hline
      \multirow{2}{*}{B7}   &Pr\ccsl$^*$& \tabincell{l}{{$ctrlInDe30\ \triangleq\ ctrlIn\ (30)\ \textcolor{red}{\rightsquigarrow}\ ms$}\\
      $ctrlInDe100\ \triangleq\ ctrlIn\ (100)\ \textcolor{red}{\rightsquigarrow}\ ms$\\
      {$\textcolor{red}{\preceq_{0.95}}(ctrlInDe30, ctrlOut, ctrlInDe100) $}} &140 & valid & 8.82 & 8.4 &25.31 \\\cline{2-3}
    & UPPAAL & \tabincell{l}{Pr[$\leqslant$10000]([ ] forall (i:int[1,2]) $\neg B7\_Causality(i).fail$) $\geq 0.95$} & & & & &\\
   \hline
    \end{tabular}%
  \label{table:cas}%
\end{table*}%

The specifications and verification results of B1--B7 are shown in Table \ref{table:cas}.
In the Pr\ccsl$^*$ specification of B1, $runAtXDir$ indicates that the vehicles are running at the positive x-direction (``$direction$=$xDir$'').
$overTake$ represents the event that the position of follow on x-axis is greater than that of lead vehicle ($x_1\geq x_0$).
B1 limits that $runAtXDir$ and $overTake$ can not happen at the same time, which can be expressed by {\small\gt{exclusion}} \emph{relation}.
In the specification of B2, $leadBrake$ ($followBrake$) denotes the event that the lead (follow) vehicle starts to brake.
$brakeDelay500$ is built by delaying $leadBrake$ for 500ms.
Thus B2 can be specified as a {\small \gt{precedence}} \emph{relation} between $followBrake$ and $brakeDelay500$.
Similarly, B3--B4 and B6--B7 can be specified. To specify B5, a periodic clock $prdClk$ that ticks every 30 ms is generated by using {\small\gt{PeriodicOn}} \emph{expression}. $leadSpeedTrig$ and $followSpeedTrig$ represent the triggering events of speed sensors of the lead and follow vehicles. B5 is specified as a ternary {\small\gt{coincidence}} \emph{relation}.

\begin{figure}[htbp]
  \centering
  \includegraphics[width=2.5in]{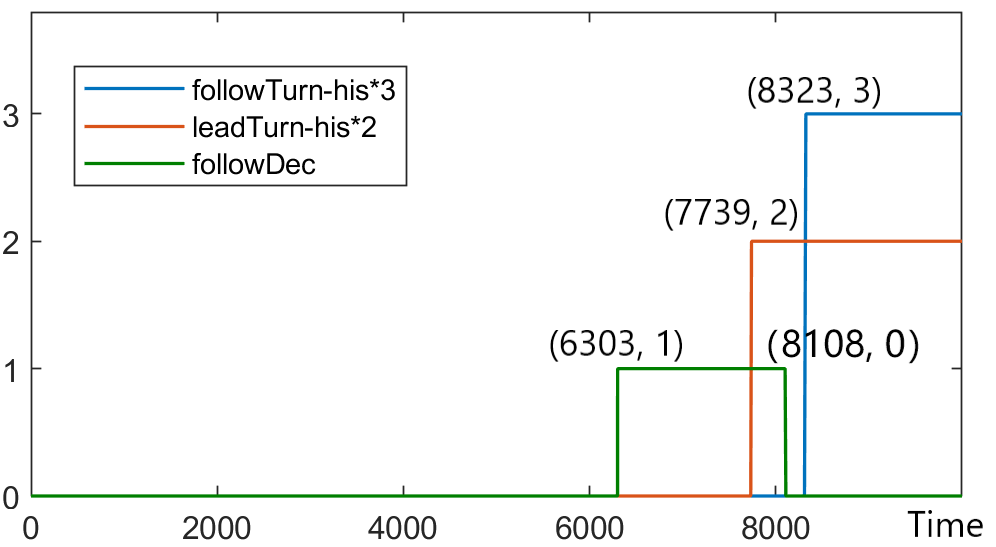}
  \caption{CE of B4: \emph{leadTurn-his} (\emph{followTurn-his}) represents the history of the clock (event) that the lead (follow) vehicle turns left.  At Time=7739, the lead vehicle starts to turn left (\emph{leadTurn-his} becomes 1), while the follow vehicle is under deceleration mode (represented by ``\emph{followDec}==1'').  The follow does not turn until it finishes the deceleration at Time=8108 and starts to turn left at Time=8323 (\emph{followTurn-his} becomes 1), which violates B4.
  }
  \label{fig:CE}
\end{figure}

The verification results in Table \ref{table:cas} shows that B1--B3 and B5--B7 are established as valid while B4 is unsatisfied.
The invalid property B4 is identified using ProTL which generates a counter-example (CE) presented in Fig. \ref{fig:CE}.
 After analyzing CE, the cause of error was found: when the follow vehicle is decreasing its speed and the lead vehicle turns left, the follower keeps speeding down but does not turn left until the deceleration is completed. Based on the CE, the system model are refined: when the follower is under deceleration mode and it detects that the lead vehicle turns left, the follower first turns left and then continues to speed down after turning.
After the modification, B4 becomes valid.

\section{Related Work}
\label{sec:r-work}
In the context of \ed, efforts on the integration of \ed\ and formal techniques were investigated in several works  \cite{goknil2013tool, ksafecomp11, abs-ccsl, lei2010, kiceccs13}, which are however, limited to the executional aspects of system functions without addressing dynamic and stochastic behaviors. Kang \cite{ksac14} defined the execution semantics of both the controller and the environment of industrial systems in \ccsl\ which are also given as mapping to \uppaal\ models amenable to model checking.
Du et al. \cite{du2018pcssl} proposed a probabilistic extension of \ccsl, called p\ccsl, for specifying the stochastic behaviors of uncertain events in cyber-physical (CPS) and provided the transformation from p\ccsl\ into Stochastic Hybrid Automata.
In contrast to our current work, those approaches lack precise annotations specifying continuous dynamic behaviors in particular regarding different clock rates during execution.

Transformation of \ccsl\ specifications into verifiable formalisms for formal analysis has been investigated in several works \cite{Yin2011Verification, chen2018model}. Yin et al. \cite{Yin2011Verification} translated \ccsl\ specifications into Promela models amenable to model checking using SPIN model checker.  Chen et al. \cite{chen2018model} performed formal analysis of timed behaviors specified in \ccsl\ by mapping the specifications into timed Input/Output automata. However, neither tool support for automatic transformation nor probabilistic analysis is provided in those works.
Zhang et al. \cite{Zhang2016An} implemented a tool {\small \gt{clyzer}} for formal analysis of \ccsl\ constraints through automated translation from \ccsl\ specifications into SMT formulas amenable to SMT solving.
Compared to their approach, we provide the tool support for probabilistic analysis of dynamic and stochastic systems behaviors based on the translation from Pr\ccsl$^*$ specifications into formal models.

\section{Conclusion}

\label{sec:conclusion}
In this paper, we present a tool-supported approach for formal verification of dynamic and stochastic behaviors for automotive systems.
To enable the formal specifications of stochastic behaviors and continuous dynamics in automotive systems,
we propose an extension of Pr\ccsl, i.e., Pr\ccsl$^*$, which is augmented with notations
for descriptions of continuous/discrete variations of physical quantities, stochastic time delays, activations of actions and nondeterministic alternatives. Moreover, to support the specification of complex requirements involved with multiple events, we extend the binary \emph{relations} into n-ary \emph{relations} in Pr\ccsl$^*$.
To enable the formal verification of system behaviors/requirements specified in Pr\ccsl$^*$, we provide the mapping rules to translate Pr\ccsl$^*$ specifications into verifiable \smc\ models.
Based on the proposed translation strategies, we implement an automatic translation tool, namely \protl, which also supports verification of the translated models by leveraging \smc\ as an analysis backend.
The applicability of our approach and tool is demonstrated by conducting verification of (non)-functional properties on two automotive system case studies.

As ongoing work,  formal validation of the correctness of translation rules from Pr\ccsl$^*$ into stochastic \smc\ models is further investigated. Furthermore, new features of \protl\ with respect to analysis of \smc\ models involving
wider range of variable/query types (e.g., \emph{urgent channels} and \emph{bounded integers}) are further developed.

\vspace{0.1in}
\noindent\textbf{Acknowledgment.} This work is supported by the EASY project funded by NSFC,
a collaborative research between Sun Yat-Sen University and University of Southern
Denmark.

\bibliographystyle{IEEEtran}
\bibliography{reference}

\end{document}